\documentclass[a4paper,runningheads]{llncs}
\usepackage[T1]{fontenc}
\usepackage{amsmath,amssymb,mathtools}
\usepackage{booktabs,tabularx,array}
\usepackage{algorithm,algpseudocode,float,flafter,placeins}
\usepackage{graphicx,tikz,pgfplots}
\usetikzlibrary{arrows.meta,positioning,fit,calc,backgrounds}
\pgfplotsset{compat=1.18}
\usepgfplotslibrary{groupplots,fillbetween}
\usepackage{microtype}
\usepackage[hidelinks]{hyperref}
\definecolor{paperblue}{RGB}{37,75,96}
\definecolor{paperred}{RGB}{156,48,55}
\definecolor{papergreen}{RGB}{34,111,90}
\numberwithin{equation}{section}
\newcommand{\Gen}{\mathcal L}
\newcommand{\Adm}{\operatorname{Adm}}

\newcommand{\J}{J_{\mathrm{DPO}}}
\newcommand{\sem}[1]{\langle\!\langle #1\rangle\!\rangle}
\newcolumntype{Y}{>{\raggedright\arraybackslash}X}
\DeclareRobustCommand{\NoEdge}{\ifmmode\text{\texttt{NoEdge}}\else\texttt{NoEdge}\fi}
\newcommand{\PaperGuideURL}{https://github.com/dmerkle/molfoundry-observables/blob/ce8f37fd2b0466fbaf9c44098cdc37c175da550a/README.md}
\newcommand{\SourceRoot}{https://github.com/dmerkle/molfoundry-observables/blob/ce8f37fd2b0466fbaf9c44098cdc37c175da550a}

\definecolor{codelinkblue}{HTML}{005BBB}
\DeclareRobustCommand{\algsource}[3]{\href{\SourceRoot/#1\#L#2}{\textcolor{codelinkblue}{#3\,\ensuremath{\nearrow}}}}
\algdef{SE}[CODEWHILE]{CodeWhile}{EndCodeWhile}[3]{\algsource{#1}{#2}{\algorithmicwhile\ #3\ \algorithmicdo}}{\algorithmicend\ \algorithmicwhile}
\algdef{SE}[CODEFOR]{CodeFor}{EndCodeFor}[3]{\algsource{#1}{#2}{\algorithmicfor\ #3\ \algorithmicdo}}{\algorithmicend\ \algorithmicfor}
\newcommand{\MaterialsURL}{https://github.com/dmerkle/molfoundry-observables/blob/ce8f37fd2b0466fbaf9c44098cdc37c175da550a/docs/paper-claims.md}

\newcommand{\ChlorinationResultsURL}{https://github.com/dmerkle/molfoundry-observables/blob/ce8f37fd2b0466fbaf9c44098cdc37c175da550a/scripts/reproduction/README.md}
\newcommand{\chlorinationdetails}{\appref{app:chlorination-equations}{B.6}}

\newcommand{\ExamplePagesURL}{https://stochmech.algochem.techfak.de/}
\definecolor{weblinkteal}{HTML}{006B65}
\DeclareRobustCommand{\examplepage}[2]{\begingroup\setlength{\fboxsep}{1.5pt}\href{\ExamplePagesURL experiments/#1/index.html}{\fcolorbox{weblinkteal}{weblinkteal!5}{\textcolor{weblinkteal}{\footnotesize\sffamily #2\,\ensuremath{\nearrow}}}}\endgroup}
\newcommand{\ExampleNotebookRoot}{https://stochmech.algochem.techfak.de/notebooks/lab/tree/notebooks/examples}
\DeclareRobustCommand{\examplenotebook}[2]{\begingroup\setlength{\fboxsep}{1.5pt}\href{\ExampleNotebookRoot/#1.ipynb}{\fcolorbox{codelinkblue}{codelinkblue!5}{\textcolor{codelinkblue}{\footnotesize\sffamily #2\,\ensuremath{\nearrow}}}}\endgroup}
\DeclareRobustCommand{\examplelinks}[3]{\examplepage{#1}{#3 webpage}\space\examplenotebook{#2}{#3 notebook}}
\newcommand{\mainref}[1]{\ref{#1}}
\newcommand{\appref}[2]{Appendix~\ref{#1}}
 \hypersetup{pdftitle={Automating Observable Closure Analysis for Graph Transformation Models with MolFoundry},pdfauthor={Marcel Friedrichs; Addie Jordon; Juri Kolčák; Daniel Merkle}}
\begin{document}
\title{Automating Observable Closure Analysis\\for Graph Transformation Models\\with MolFoundry}
\titlerunning{Observable Closure Analysis with MolFoundry}
\author{Marcel Friedrichs\inst{1,2} \and
Addie Jordon\inst{1,2} \and
Juri Kol\v c\'ak\inst{1,2} \and
Daniel Merkle\inst{1,2,3}}
\authorrunning{M. Friedrichs et al.}
\institute{Algorithmic Cheminformatics Group, Faculty of Technology,\\
Bielefeld University, Germany
\and
Center for Biotechnology (CeBiTec), Bielefeld University, Germany
\and
Department of Mathematics and Computer Science (IMADA),\\
University of Southern Denmark, Odense, Denmark}
\maketitle
\begin{abstract}
Graph transformations model changing systems, with applications in software,
biology and chemistry, among others.
We introduce MolFoundry, a graph transformation framework, and present a
workflow for observable analysis that tracks the dynamics of local patterns
through rule composition. Analysing rule compositions avoids the need to
enumerate a potentially infinite state space.
Building on rule-algebraic stochastic rewriting, MolFoundry's kernel composes rules
with nested application conditions, including missing-edge constraints.
Given rules, rates and query patterns, the algorithms iteratively identify
auxiliary pattern counts and search for closure. Using exact counting
identities supplied by graph-family backends, successful searches yield
closed ordinary differential equations for average counts or, with justified
individual jumps, finite event models describing joint changes and rates.
We evaluate conditional composition and automated discovery of auxiliary
observables against published calculations and independent rule application.
Applications to aldol addition, copolymerization and chlorination show
how the framework yields quantitative predictions directly from molecular
graph rules.
\end{abstract}

\section{Introduction}
\label{sec:introduction}

Graph transformations are an established approach to modelling systems whose
structures change. Applications range from software models and visual
languages~\cite{Taentzer2004AGG,GhamarianEtAl2012GROOVE} to chemical reaction
systems~\cite{AndersenFlammMerkleStadler2016MOD} and biological interaction
networks~\cite{BoutillierEtAl2018Kappa}. Graphs represent entities and their
relationships; local rules specify how these entities and relationships can be created, deleted or
reconnected. A key strength is that this graphical description has rigorous mathematical semantics. In algebraic graph transformation, rules themselves can be composed, providing a formal connection between individual
transformations and their combined
effects~\cite{BehrHarmerKrivine2023,BehrKrivine2021}.

Several tools make substantial parts of this framework available in practice. AGG supports modelling, execution and validation of graph transformation systems~\cite{Taentzer2004AGG}; GROOVE provides state-space
exploration and model checking~\cite{GhamarianEtAl2012GROOVE}.
M\O D supports molecular graph rewriting, reaction-network generation and
rule composition~\cite{AndersenFlammMerkleStadler2016MOD}, while the Kappa
platform provides simulation and analysis of biochemical
models~\cite{BoutillierEtAl2018Kappa}.
Despite the range of graph transformation tools, not every theoretical capability is available as an automated analysis.

One particularly compelling capability is to infer quantitative behaviour
directly from rules, their rates and a few query patterns. How does the count
of each pattern change over time, on average? Which counts change together
and which counts influence others? These count functions are called \emph{observables}; expectations of their values,
products and powers are their \emph{moments}. Behr and collaborators developed
a rule-algebraic calculus for their evolution and, through polynomial
jump closure, for converting joint moment dynamics into equations for
generating functions~\cite{BehrDanosGarnier2020}. This permits the analysis of
selected counts without enumerating every graph state, even when the
rules generate a potentially infinite state space. Conditional composition
theory~\cite{BehrKrivine2021}, and its application
to stochastic models in the life sciences~\cite{BehrKrivineAndersenMerkle2021},
extend these foundations to rules and observables with structural
conditions, such as requiring a connection to be absent (a missing-edge constraint).

Important steps toward executable analysis already exist. FRAGGER
automates the expansion of graph rate equations and recursively introduces
the counts needed to express their
changes~\cite{DanosHeindelHonoratoZimmerStucki2020}; KaDE compiles Kappa
rules into reaction-network ordinary differential equation (ODE) models with symmetry-based
reductions~\cite{CamporesiFeretLy2017KaDE}. Earlier Z3 prototypes address
constrained overlaps for rule composition and commutator
calculations~\cite{BehrHeckelGhaffariSaadat2020}.
These are substantial building blocks, but the published implementations
do not combine conditional composition, automatic discovery of auxiliary
observables, and extraction of their joint jumps and rates.
FRAGGER identifies nested application conditions as future
work~\cite[Section 4]{DanosHeindelHonoratoZimmerStucki2020}.
In a published molecular case study, a conditioned observable had to be
replaced by an unconditioned one because the implementation could not
represent its condition~\cite[Section 8.3.1]{BehrKrivineAndersenMerkle2021}.

Related reductions include stochastic lumping of rule-based
models~\cite{FeretHenzingerKoepplPetrov2012} and constrained lumping of
existing ODE systems, as in CLUE~\cite{OvchinnikovEtAl2021CLUE}.
Unlike reduction of an existing ODE system, our analysis discovers count
dynamics directly from conditional graph rules, even when no finite
reaction network is available.

Observable analysis repeatedly computes how each rule changes a selected
pattern count. Conditional composition gives an expression for its
instantaneous expected change~\cite{BehrKrivineAndersenMerkle2021}.
New pattern counts in the resulting expression are analysed in turn, forming an
iterative dependency-discovery process.

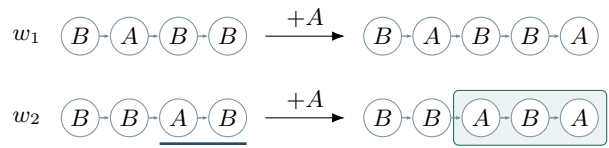
\begin{figure}[!tb]
\centering
\begin{tikzpicture}[x=1cm,y=.9cm,font=\small,>=Latex,
  bead/.style={circle,draw=paperblue!60,minimum size=5mm,inner sep=0pt},
  edge/.style={-{Latex[length=2pt]},line width=.5pt,paperblue!65}]
\node[anchor=east] at (-.4,0) {$w_1$};
\node[anchor=east] at (-.4,-1.2) {$w_2$};
\foreach \row/\seq in {0/{B,A,B,B},1/{B,B,A,B}} {
  \foreach \a [count=\i from 0] in \seq {
    \node[bead,fill=white] (l\row\i) at (.65*\i,-1.2*\row) {$\a$};
    \ifnum\i>0\pgfmathtruncatemacro{\p}{\i-1}
      \draw[edge] (l\row\p)--(l\row\i);\fi
  }
}
\foreach \row/\seq in {0/{B,A,B,B,A},1/{B,B,A,B,A}} {
  \foreach \a [count=\i from 0] in \seq {
    \node[bead,fill=white] (r\row\i) at (4+.65*\i,-1.2*\row) {$\a$};
    \ifnum\i>0\pgfmathtruncatemacro{\p}{\i-1}
      \draw[edge] (r\row\p)--(r\row\i);\fi
  }
  \draw[->] (2.45,-1.2*\row)--node[above,font=\footnotesize]{$+A$}(3.5,-1.2*\row);
}
\draw[paperblue,line width=1pt] (1.05,-1.58)--(2.2,-1.58);
\begin{scope}[on background layer]
\node[fit=(r12)(r13)(r14),draw=papergreen,fill=papergreen!8,
  rounded corners=2pt,inner sep=3pt] {};
\end{scope}
\end{tikzpicture}
\caption{The paths $w_1=BABB$ and $w_2=BBAB$ both have zero $ABA$
occurrences. Appending $A$ leaves the upper count unchanged but creates
one occurrence below (green), because only $w_2$ ends in $AB$ (blue).
This ending count is an auxiliary observable discovered by MolFoundry.}
\label{fig:suffix}
\end{figure}
 
\begin{samepage}
The need for auxiliary observables is already visible in a growing A/B chain.
The two paths in Figure~\ref{fig:suffix} both have zero occurrences of
$ABA$, but appending $A$ creates an $ABA$ occurrence only for the path
ending in $AB$. The change in the $ABA$ count therefore depends on the
$AB$ ending count; the motif count alone cannot describe its own changes.
This running example follows the analysis in
Section~\ref{sec:tool} and later corresponds to a copolymer sequence.

\end{samepage}

A successful search can yield a finite closed system of ODEs for expected counts~\cite{DanosHeindelHonoratoZimmerStucki2020},
or a finite event model recording simultaneous count changes and rates.
The latter retains fluctuations and dependencies beyond the means, but
need not yield a finite expectation ODE. Polynomial expressions for average
changes alone do not determine individual jumps. Neither kind of closure
is guaranteed for arbitrary rules and queries~\cite[Section~6]{BehrDanosGarnier2020}.

We present an executable workflow for observable analysis built on
MolFoundry, combining
multiplicity-preserving conditional composition, sound graph-family pruning,
and shared searches for exact observable dynamics.
MolFoundry's kernel computes conditional rule compositions, including
missing-edge tests (\NoEdge{}) and transport of nested application
conditions.
On top of this core, shared algorithms search for linear expectation
closure or discover polynomial support and, with justified individual
jumps, construct finite event models.
Graph-family backends supply exact counting
identities; the algorithms discover which auxiliary observables a query
needs. The resulting guarantees are relative to these supplied identities
and jump arguments, not to arbitrary graph families. Unfinished searches
retain partial results and pending work.

We apply this workflow to chemistry, where structural conditions can
change quantitative predictions. We automate the conditional compositions
in the published aldol-addition case study~\cite{BehrKrivineAndersenMerkle2021}
with the observable's missing-edge
condition intact, derive sequence-count dynamics directly from atom-resolved
copolymerization rules on chains of unbounded size, and analyse a stylised
chlorination model in which a negative application condition
(NAC)---a special case of nested conditions forbidding a pattern
extension---prevents further substitution.

Section~\ref{sec:foundations} states the mathematical requirements,
Section~\ref{sec:tool} describes the workflow, and
Section~\ref{sec:results} evaluates composition, query reuse and agreement
with independent calculations. The mathematical appendices and
\href{\MaterialsURL}{material guide} give counting arguments and executable
inputs.
\FloatBarrier
 \section{Background}
\label{sec:foundations}

The analysis computes how rules change counts, then seeks a finite
description of their dynamics. We recall the required conditional
double-pushout (DPO) calculus~\cite{BehrKrivine2021,BehrKrivineAndersenMerkle2021}
and distinguish the two kinds of output.

\subsection{Rules, Queries and Count Changes}

A linear rule $r=(L\leftarrow K\rightarrow R,c)$ has input $L$, output $R$,
preserved part $K$, injective maps and an application condition $c$.
A match $m:L\hookrightarrow G$ is admissible if it satisfies $c$ and the
DPO gluing conditions; $r_m(G)$ is its successor. Deleting a vertex requires
all its incident edges to be matched and
deleted~\cite[Definitions 4--5]{BehrKrivineAndersenMerkle2021}.
Our observable analysis supports conjunctions of \NoEdge$(u,v)$, forbidding
an edge between matched vertices, and local neighbour-count predicates.
Their nested-condition encodings are in \appref{app:conditions}{A.1}.
The chlorination example additionally handles a neighbour NAC using a
dedicated local reduction and joint-jump certificate.

Fix a possibly infinite family $\Omega$ of finite labelled graphs that is
closed under admissible rule applications. Conditions select matches;
$\Omega$ restricts whole states. An observable is a real-valued function
on $\Omega$; a query $(P,c)$ counts injective matches satisfying $c$,
written $q(G)=[P,c](G)$. Each admissible rule match has rate
$\kappa_r\nu_r$, with rate parameter $\kappa_r\geq0$ and rational
normalization $\nu_r>0$.

\begin{samepage}
The stochastic generator sums each possible
change in an observable $F$, multiplied by its rate:
\begin{equation}
 (\Gen F)(G)=\sum_r\kappa_r\nu_r
   \sum_{m\in\Adm_r(G)}[F(r_m(G))-F(G)].
 \label{eq:generator}
\end{equation}
\end{samepage}
Here $\Adm_r(G)$ is the admissible-match set and
$\Gen=\sum_r\kappa_r\Gen_r$ defines rate-free rule contributions.
Rules sharing a parameter form a rate channel $\gamma$, with
$\kappa_r=\kappa_\gamma$ and $\Gen_\gamma=\sum_{r\in\gamma}\Gen_r$;
independent rates use singleton channels. Write $G_t$ for the associated
continuous-time Markov chain (CTMC).

To compute a query's changes, represent it by the identity rule
$O_i=(P_i\leftarrow P_i\rightarrow P_i,c_i)$.
These queries are conditional pattern counts; general DPO
observables can also use a proper interface $K\subsetneq P_i$~\cite[Theorem~6]{BehrKrivineAndersenMerkle2021}.
With $r_1$ applied first, conditional composition
$r_2\star r_1=\sum_\mu r_2\circ_\mu r_1$
sums over admissible overlaps between the first rule's output and the
second's input, retaining both conditions~\cite{BehrKrivine2021}.
Multiplicities of isomorphic composites are retained as coefficients in
finite formal rational linear combinations of rule basis elements.
Projection
$\J(r)=(L\leftarrow K\rightarrow L,c)$ preserves $K$ and admissibility.
We extend $\J$ linearly to rule expressions.
The two composition orders count occurrences after and before a rule
application; unchanged occurrences cancel. The commutator gives the
normalized count change~\cite[Theorems 6 and 9, Corollary 1]{BehrKrivineAndersenMerkle2021}:
\begin{equation}
 \Gen_rq_i=\nu_r\sem{\J(O_i\star r-r\star O_i)}\quad\text{on }\Omega.
 \label{eq:count-change}
\end{equation}
Here $\sem{\cdot}$ sums projected admissible-match counts with their
coefficients, and $\Gen_rq_i$ is the query's \emph{generator image}.
Exact count reduction rewrites this expression using identities valid
throughout $\Omega$.

\subsection{Finite Descriptions of Observable Dynamics}

\paragraph{Linear expectation closure.}
If the selected counts and all images $\Gen_\gamma b_i$ are
constant-coefficient linear combinations of a finite vector
$B=(b_1,\ldots,b_n)^{\mathsf T}$, its averages satisfy a closed ordinary
differential equation (ODE)~\cite{DanosHeindelHonoratoZimmerStucki2020}:
\begin{equation}
 \Gen_\gamma B=A_\gamma B\quad\text{on }\Omega,
 \qquad
 \frac{d}{dt}\mathbb E[B(G_t)]
 =\Bigl(\sum_\gamma\kappa_\gamma A_\gamma\Bigr)\mathbb E[B(G_t)].
 \label{eq:linear}
\end{equation}
The entries of $B$ may include products or linear combinations of counts;
$(A_\gamma)_{ij}$ is the coefficient of $b_j$ in $\Gen_\gamma b_i$.

\paragraph{Finite event model.}
For a vector $\phi$ of selected and additional pattern counts, an event
records a constant integer change $\delta_e$ and polynomial activity
$a_e$. Its rate is $\kappa_{\gamma(e)}a_e(z)$ at $z=\phi(G)$.
For real-valued functions $f$ of these coordinates, the event generator
$\overline\Gen$ must preserve the simultaneous changes:
\begin{equation}
 \begin{aligned}
 (\overline\Gen f)(z)&=\sum_e\kappa_{\gamma(e)}a_e(z)
                         [f(z+\delta_e)-f(z)],\\
 \Gen(f\circ\phi)&=(\overline\Gen f)\circ\phi.
 \end{aligned}
 \label{eq:events}
\end{equation}
Each event groups applications in one rate channel with the same jump;
activities include match normalization, are nonnegative on $\phi(\Omega)$,
and permit only jumps remaining in that set. This is a sufficient form of
polynomial jump closure~\cite[Definition 4.1]{BehrDanosGarnier2020}.
Choosing $f(z)=z_i$ gives
$\overline\Gen z_i=\sum_e\kappa_{\gamma(e)}\delta_{e,i}a_e(z)$.

A finite event model determines the exact expectation equations, but
polynomial activities may introduce an infinite moment hierarchy even if finite linear closure exists.
Conversely, opposite jumps may cancel in the mean drift, yielding a small exact expectation system even when
these coordinates admit no finite constant-jump event model.
Neither output is universally preferable.
Linear closure therefore remains a separate reduction problem even after event closure succeeds.

For the process interpretation we assume nonexplosion and, for expectation
equations, the required integrability and interchange with the generator.
\appref{app:emgf}{A.2} gives these qualifications and the conversion from
events to a joint exponential moment-generating function (EMGF) equation.
 \section{Observable Analysis with MolFoundry}
\label{sec:tool}

Within the rule and query forms supported by a backend, users supply
rules, rates, query patterns and search budgets. The backend provides
reusable counting identities and the justification required for event
construction, not the query's closure basis or equations.
MolFoundry's kernel computes the rule compositions in
Equation~\eqref{eq:count-change};
its Python analysis package \texttt{mfobs} discovers auxiliary observables
and searches for closed dynamics. Changing the query reuses the backend;
a new graph family may require additional counting and jump arguments.

For the running $ABA$ query, discovery introduces free-token and
path-ending counts and constructs eight joint event classes.
Section~\ref{sec:path-backend} works through this result after the two
search procedures below.

\subsection{Conditional Composition and the Counting Interface}
\label{sec:conditional-kernel}

The two algorithms computing the linear and event closures share composition and reduction backends.

The composition kernel enumerates common overlaps, transports application
conditions to composite inputs and aggregates their multiplicities.
It also implements generic nested-condition transport, subject to a
budget on the enumeration used to transport conditions; the observable
analysis uses the condition fragment stated in Section~\ref{sec:foundations}.
The analysis composes each query identity with each source rule in both
orders and applies jump projection as in Equation~\eqref{eq:count-change}.
Terms are collected by a canonical key for the span and its represented
conditions; span isomorphism alone is insufficient. This comparison does not
decide logical equivalence of arbitrary nested conditions, which is
undecidable in general~\cite{HabelPennemann2009}, but suffices for the
conditions used in our examples. Further count equalities on $\Omega$ are
established by the counting backend.

The counting backend $\mathcal A$ implements exact reductions on the
specified graph family $\Omega$. Its operation $\mathcal A.\mathsf{Reduce}$
rewrites a projected rule expression from the composition kernel in terms of pattern counts, e.g. for the $A_\gamma$ matrix from Equation~\eqref{eq:linear}.
It returns the resulting expression, the patterns whose counts occur in it, and the
identities used. A \emph{pattern factor} is one of these count functions
appearing as a factor in a product. A term can be removed as zero only when
its count vanishes throughout $\Omega$. Products likewise require a
counting identity valid on $\Omega$.
Reduction preserves equality on $\Omega$ but need not recognize every
identity. Exact elimination in the backend's representation can therefore
leave redundant count functions, as in the eight-site binding model
(Section~\ref{sec:eight-sites}).

In the growing A/B path model (Section~\ref{sec:path-backend}), a pattern
combining a free $A$ token with a word $w$ anchored
at the unique path end has $M_AS_w$ matches: $M_A$ counts free tokens and
$S_w$ indicates the ending. Their matches are necessarily disjoint, so the
counts multiply. This identity holds for arbitrary finite paths and token
pools. The backend implements it for supported words; composition and the
worklist determine which words the selected query requires.

The number of overlap candidates in the composition kernel can be reduced by bounds on incident edges in the graph family $\Omega$.
For molecular graphs, these bounds constrain how
many bonds of each type may connect an atom to neighbours of specified
labels. These supplied restrictions are not a general chemical-validity check.
Further identifications may merge vertices and
edges that are still distinct in a partial overlap. The enumerator
therefore discards all extensions of a partial overlap only if even the
largest possible further merging cannot satisfy a supplied bound. Each
remaining candidate still undergoes conditional composition.
\appref{app:examples}{B} gives the bounds and counting arguments
used by the supported backends.

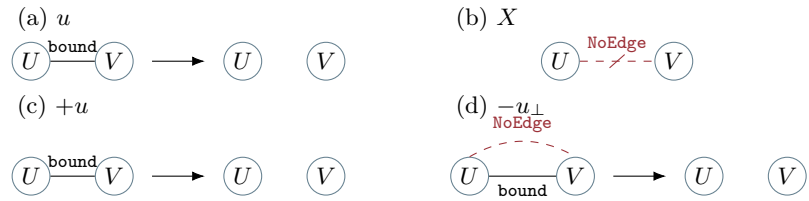
\begin{figure}[!tb]
\centering
\begin{tikzpicture}[x=1cm,y=.8cm,font=\small,>=Latex,
  vertex/.style={circle,draw=paperblue!70,fill=white,minimum size=5mm,inner sep=0pt}]
\node[anchor=west] at (-.3,1.6) {(a) $u$};
\node[vertex] (ul) at (0,.9) {$U$};
\node[vertex] (vl) at (1.1,.9) {$V$};
\draw (ul)--node[above,font=\scriptsize]{\texttt{bound}}(vl);
\draw[->] (1.6,.9)--(2.3,.9);
\node[vertex] at (2.8,.9) {$U$};
\node[vertex] at (3.9,.9) {$V$};
\node[anchor=west] at (5.5,1.6) {(b) $X$};
\node[vertex] (ux) at (7,.9) {$U$};
\node[vertex] (vx) at (8.5,.9) {$V$};
\draw[dashed,paperred] (ux)--node[above,font=\scriptsize]{\NoEdge{}}(vx);
\draw[paperred] (7.65,.79)--(7.85,1.01);

\node[anchor=west] at (-.3,.15) {(c) $+u$};
\node[vertex] (up) at (0,-1) {$U$};
\node[vertex] (vp) at (1.1,-1) {$V$};
\draw (up)--node[above,font=\scriptsize]{\texttt{bound}}(vp);
\draw[->] (1.6,-1)--(2.3,-1);
\node[vertex] at (2.8,-1) {$U$};
\node[vertex] at (3.9,-1) {$V$};

\node[anchor=west] at (5.5,.15) {(d) $-u_\bot$};
\node[vertex] (un) at (5.8,-1) {$U$};
\node[vertex] (vn) at (7.2,-1) {$V$};
\draw (un)--node[below,font=\scriptsize]{\texttt{bound}}(vn);
\draw[dashed,paperred] (un.north) to[bend left=30]
  node[above,font=\scriptsize]{\NoEdge{}} (vn.north);
\draw[->] (7.7,-1)--(8.4,-1);
\node[vertex] at (8.9,-1) {$U$};
\node[vertex] at (10,-1) {$V$};
\end{tikzpicture}
\caption{Conditions in the two composition orders. (a) Rule $u$ deletes the
\texttt{bound} edge, preserving both vertices. (b) Query $X$ counts pairs
without that edge (red dashed line). (c) Full overlap in $O_X\star u$ gives
$+u$: deletion establishes absence. (d) The reverse order gives $-u_\bot$:
its input requires the edge both present and absent, so the condition is
unsatisfiable.}
\label{fig:conditional-composition}
\end{figure}

\paragraph{Illustration: counting a missing edge.}
Let $u$ delete a \texttt{bound} edge between differently labelled vertices
$U,V$. Let $X$ count such pairs with no edge, represented by the identity
rule $O_X$. For the count reduction, take the two-state family containing
exactly one $U,V$ pair and only the optional \texttt{bound} edge.
In the full overlap of $O_X\star u$, deletion establishes the
query's absence condition, leaving a positive contribution from $u$.
For this overlap, the reverse order $u\star O_X$ requires the input edge
to be both present and absent, giving a contradictory composite $u_\bot$
(Figure~\ref{fig:conditional-composition}).

The one-endpoint overlaps cancel between the two orders, as does the empty
overlap when included. Projection leaves the bound-edge count, which is
$1-X$ on this two-state family. With unit match normalization,
rule $u$ contributes $\kappa_u(1-X)$ to $\Gen X$.
\appref{app:conditions}{A.1} gives the full overlap expression.

\subsection{Discovering Observable Dynamics}
\label{sec:linear-discovery}

Algorithm~\ref{alg:linear} applies the observable-dependency expansion
underlying graph rate equations~\cite[Section 1.2]{DanosHeindelHonoratoZimmerStucki2020}
to the conditional count-change calculus~\cite{BehrKrivineAndersenMerkle2021}.
It seeks a vector $B$ satisfying Equation~\eqref{eq:linear}, starting with
the count observables of the query set $Q$. A worklist $W$ holds functions
whose generator images still need to be computed. For each function and
rate channel, exact elimination attempts to express its image in the current
entries of $B$, using the backend's representation. If it succeeds, the
algorithm stores the coefficients $v_j$, which form a row of the $A_\gamma$ matrix for each rate channel.
Otherwise, it adds the residual expression and queues that function. Entries of $B$ may be pattern counts,
products or linear combinations. Contributions from rules sharing a rate
parameter are summed before this test; one nonzero residual adds one
coordinate.

\begin{algorithm}[!tb]
\caption{Linear expectation-closure search.}
\label{alg:linear}
\begin{algorithmic}[1]
\Require Rules $\mathcal R$ grouped by rate channel, queries $Q$, backend $\mathcal A$ for $\Omega$, budgets
\State \algsource{src/mfobs/core/linear_closure.py}{95}{$B\gets$ count observables of $Q$; $W\gets B$} \label{line:closure-input}
\CodeWhile{src/mfobs/core/linear_closure.py}{114}{$W\ne\varnothing$ and budget remains}
 \State \algsource{src/mfobs/core/linear_closure.py}{115}{Remove $b_i$ from $W$} \label{line:closure-pop}
 \CodeFor{src/mfobs/core/linear_closure.py}{120}{each rate channel $\gamma$}
  \State \algsource{src/mfobs/core/linear_closure.py}{122}{Compute $F\gets\sum_{r\in\gamma}\Gen_rb_i$ by \textup{(\ref*{eq:count-change})}} \label{line:closure-image}
  \State \algsource{src/mfobs/core/linear_closure.py}{125}{Reduce $F$ modulo $\operatorname{span}(B)$ in $\mathcal A$; add its nonzero residual} \label{line:closure-reduce}
  \State \algsource{src/mfobs/core/linear_closure.py}{158}{Queue the new function, if any; store the exact coefficients of $F$} \label{line:closure-extend}
 \EndCodeFor
\EndCodeWhile
\State \Return matrices $A_\gamma$ if complete; otherwise partial equations and pending work \label{line:closure-return}
\end{algorithmic}
\end{algorithm}

\label{sec:event-discovery}
Algorithm~\ref{alg:main} seeks a finite event model in two stages:
polynomial-support discovery and event extraction. Starting from the
query counts $\phi$, it reduces each rule's activity
$a_r(G)=\nu_r|\Adm_r(G)|$ and each generator image
$F_{i,r}=\Gen_r\phi_i$ to polynomials in these counts. A new pattern
factor becomes a coordinate of $\phi$, and its identity rule enters $W$.
The record $T$ stores the compositions and reductions. When no new
pattern factors are needed, all activities and coordinate images are
polynomials in the discovered vector $\phi$; this is
\emph{polynomial support closure}.

Event extraction requires \emph{uniform matches}: at a fixed host graph,
every admissible match of a rule has the same change $\Delta_r\phi(G)$.
For $a_r>0$, this change is $F_{\cdot,r}/a_r$, where $F_{\cdot,r}$ is the
vector of coordinate images. The backend partitions the enabled coordinate states
into finitely many nonempty classes on which this vector is constant.
Each polynomial indicator $I_c$ is one in its class and zero elsewhere,
so the corresponding event has activity $a_rI_c$. The set $E$ records
its rule, jump and activity. The conversion in \appref{app:emgf}{A.2} then
gives the joint moment-generating-function equation.

The growing A/B path model in Section~\ref{sec:path-backend} uses this
uniform-match loop.
R\'emy trees instead use a context-counting argument; neighbour-NAC
chlorination partitions source matches by their complete jumps using
symbolic condition reduction and locality. These separately justified
constructions satisfy Equation~\eqref{eq:events}.

\begin{algorithm}[!tb]
\caption{Pattern-support discovery and uniform-match event extraction.}
\label{alg:main}
\begin{algorithmic}[1]
\Require Rules $\mathcal R$, queries $Q$, exact backend $\mathcal A$ for $\Omega$, budgets
\State \algsource{src/mfobs/core/pattern_support.py}{126}{$\phi\gets$ count observables of $Q$; $W\gets$ their identities; initialize $T$} \label{line:event-support}
\CodeFor{src/mfobs/core/pattern_support.py}{161}{each rule $r$}
 \State \algsource{src/mfobs/core/pattern_support.py}{163}{$a_r\gets\mathcal A.\mathsf{Reduce}(\nu_r\J(r))$} \label{line:event-activity}
 \State \algsource{src/mfobs/core/pattern_support.py}{164}{Add newly occurring represented pattern factors to $\phi,W$}
\EndCodeFor
\CodeWhile{src/mfobs/core/pattern_support.py}{167}{$W\ne\varnothing$ and budget remains}
 \State \algsource{src/mfobs/core/pattern_support.py}{170}{Remove $O_i$ representing $\phi_i$ from $W$}
 \CodeFor{src/mfobs/core/pattern_support.py}{172}{each rule $r$}
  \State \algsource{src/mfobs/core/pattern_support.py}{173}{$F_{i,r}\gets\mathcal A.\mathsf{Reduce}(\nu_r\J(O_i\star r-r\star O_i))$} \label{line:event-image}
  \State Store operands, coefficients and reductions in $T$
  \State \algsource{src/mfobs/core/pattern_support.py}{174}{Add newly occurring represented pattern factors to $\phi,W$}
 \EndCodeFor
\EndCodeWhile
\If{a reduction is unsupported or work remains}
 \State \algsource{src/mfobs/core/pattern_support.py}{135}{\Return partial support, $T$ and pending work}
\EndIf
\State Require the backend's uniform-match justification \label{line:event-uniform}
\State \algsource{src/mfobs/core/uniform_events.py}{44}{$E\gets\varnothing$}
\CodeFor{src/mfobs/core/uniform_events.py}{46}{each rule $r$ with nonzero activity}
 \State \algsource{src/mfobs/core/uniform_events.py}{50}{Reduce $d_r=F_{\cdot,r}/a_r$ exactly on $a_r>0$} \label{line:event-quotient}
 \State \algsource{src/mfobs/core/uniform_events.py}{54}{Require a finite exhaustive partition $\{I_c\}$ with constant $d_r=\delta_{r,c}$} \label{line:event-refine}
 \State \algsource{src/mfobs/core/uniform_events.py}{65}{Add all $(r,\delta_{r,c},\mathcal A.\mathsf{Reduce}(a_rI_c))$ to $E$}
\EndCodeFor
\State \Return $\phi,E$, assumptions and $T$; export the EMGF equation \label{line:event-return}
\end{algorithmic}
\end{algorithm}

\paragraph{Backend-relative guarantees.}
Successful completion of Algorithm~\ref{alg:linear} establishes the
invariant-span condition~\eqref{eq:linear}.
For Algorithm~\ref{alg:main}, successful completion establishes the event
identity~\eqref{eq:events} separately for each rate channel under the
uniform-match and finite-partition requirements stated above.
Both guarantees assume sound composition and exact backend reduction
on a rule-closed graph family $\Omega$.

Both algorithms run independently from the supplied rules and queries.
Their worklists and uniform-match event extraction are shared; counting
identities and partitions are supplied by the backends. The analysis records
support closure, event construction and linear expectation closure separately.
Budget exits retain pending work and do not establish non-closure;
algorithm lines link to their implementations.

\subsection{Worked Example: A Motif in a Growing A/B Path}
\label{sec:path-backend}
\label{sec:workflow}

We now specify the growing path used informally in the introduction.
A state contains one rooted, directed path whose vertices carry labels
$A$ or $B$, together with finite pools of free $A$ and $B$ tokens. A distinguished end identifies
where the path grows, including when it is empty. Two feed rules add free
tokens at rates $\beta_A,\beta_B$. Two append rules consume a free token and
extend the path by its label, with per-match rates $k_A,k_B$. The path and
pools may grow without a fixed bound.
\examplelinks{rop-family}{02_aba}{ABA}

For a word $w$, let $C_w$ count its contiguous occurrences, including
overlaps, and $S_w$ indicate whether the path ends in $w$.
The user selects only $q=C_{ABA}$.
We run both algorithms independently for this query.
On this unbounded family, no finite channel-wise linear expectation closure
exists for $q$, so Algorithm~\ref{alg:linear} cannot close.
Algorithm~\ref{alg:main} yields a finite event model; we illustrate
the latter construction below.
Algorithm~\ref{alg:main} discovers the other counts using the path and
free-token identities in \appref{app:path-equations}{B.1}.

\paragraph{Discovering the coordinates.}
Reducing the append-rule activities
(line~\ref{line:event-activity}) introduces the free-pool counts $M_A,M_B$.
The projected commutator (line~\ref{line:event-image}) for $q$ and
appending $A$ reduces to $M_AS_{AB}$,
where $S_{AB}$ is the count of an $AB$ pattern anchored at the path end.
The generator image of $S_{AB}$ under appending $B$ introduces the
$A$-ending count $S_A$. Together, the generated expressions are
\begin{equation}
\begin{aligned}
 \Gen q&=k_AM_AS_{AB},\\
 \Gen S_{AB}&=k_BM_BS_A-(k_AM_A+k_BM_B)S_{AB},\\
 \Gen S_A&=k_AM_A(1-S_A)-k_BM_BS_A,
\end{aligned}
\label{eq:discovered-aba}
\end{equation}
with $\Gen M_\sigma=\beta_\sigma-k_\sigma M_\sigma$ for $\sigma=A,B$.
The rate-free A-append image of $M_A^nS_{AB}$ is
$-M_A^{n+1}S_{AB}$ ($n\ge1$). Repeated images of $q$ thus require
arbitrarily high powers, linearly independent on the unbounded free pool;
this rules out finite channel-wise linear closure.
No further pattern counts are needed to express these activities and
images, so polynomial support closes at
$\phi=(q,M_A,M_B,S_{AB},S_A)^{\mathsf T}$.

\paragraph{From coordinates to events.}
Every append match selects the unique path end and one free token.
Different free copies remain distinct rate contributions but give the same
coordinate jump. The unique ending also justifies
$S_A^2=S_A$, $S_{AB}^2=S_{AB}$ and $S_AS_{AB}=0$.
The path ends in $A$, in $AB$, or in neither. The corresponding class
indicators are $S_A$, $S_{AB}$ and $1-S_A-S_{AB}$.
Two append rules across these three classes, plus two feeds, give eight
events. Appending $A$ to an $AB$ ending, for example, yields
\[
 \delta_e=(1,-1,0,-1,1)^{\mathsf T},\qquad a_e=M_AS_{AB}.
\]
One motif is created, one free $A$ consumed, and the ending changes to $A$.
The full rate is $k_AM_AS_{AB}$. The conversion in \appref{app:emgf}{A.2}
gives the joint EMGF operator. \appref{app:path-equations}{B.1} lists
all eight events, one operator term and a numerical update. The expectation equation for $q$ contains
$\mathbb E[M_AS_{AB}]$, so it is not closed in the five coordinate means.

Section~\ref{sec:query-reuse} varies the query while keeping these rules
and the counting backend fixed.

\paragraph{Adding one condition.}
Add the two-state $U,V$ gate from
Figure~\ref{fig:conditional-composition} and require \NoEdge$(U,V)$ in
both append rules. Closing the gate blocks extension while feeding
continues. This is an abstract control, not a chemical inhibition mechanism.
Count reduction introduces the absence indicator $X$ as a sixth coordinate
and multiplies each append activity by $X$. Opening and closing add two
events, giving ten in total. The gate does not restore finite channel-wise
linear closure; \appref{app:path-equations}{B.1} gives the events and rates.

\subsection{Implementation and Use}
\label{sec:implementation}

MolFoundry provides graph construction, rule application, reaction-network
generation, pathway analysis and stochastic simulation.\footnote{\url{https://py.molfoundry.de/docs/}}
Its TypeScript/JavaScript kernel provides matching and conditional
composition through the \texttt{molfoundry} Python interface.
Its Python analysis package \texttt{mfobs} implements the observable analysis presented here:
shared algorithms, graph-family backends and event/EMGF construction. Its public entry point
is \texttt{mfobs.analyze}; the notebooks expose the model inputs, analysis
call and numerical settings.

The interactive examples for $ABA$, chlorination and R\'emy trees show
rules, applications, state expansions and curves. Their website generators
consume exported analysis records; notebooks use the same analysis and
numerical routines. Compositions, complete equations and counting arguments
are available as details. Inputs can be changed in Python; algorithm lines
link directly to the implementation.
 \section{Evaluation}
\label{sec:results}

The examples span conditional composition, discovery and molecular modelling.
Aldol addition tests \NoEdge{}-conditioned composition and cancellation
(Section~\ref{sec:aldol-result}). The running A/B model tests query reuse
and illustrates generated dynamics through stochastic simulation and
numerical integration of probability equations (Section~\ref{sec:query-reuse}).
R\'emy trees recover a published calculation through discovery of subtree
pattern counts (Section~\ref{sec:remy}); eight sites test \NoEdge{}-guarded
binding and closure across topologies (Section~\ref{sec:eight-sites}).
The FRAGGER comparison covers DNA-walker equations and a shared composition
benchmark (Section~\ref{sec:fragger}). Atom-resolved copolymerization realises
the A/B model with molecular rules (Section~\ref{sec:molecular-applications}).
Finally, synthetic chlorination tests a neighbour NAC on five-membered rings
with arbitrarily long alkyl tails (Section~\ref{sec:chlorination}).

Counting arguments apply throughout the declared graph families;
independent rule application, published calculations and bounded host
checks test the implementations. The
\href{\MaterialsURL}{material guide} links inputs, outputs and protocols.

\subsection{Conditional Composition for the Aldol Addition}
\label{sec:aldol-result}

\begin{figure}[!tb]
\centering
\begin{tikzpicture}[x=.965cm,y=.85cm,font=\scriptsize,
 atom/.style={circle,draw,densely dotted,minimum size=4.2mm,inner sep=0pt},
 bond/.style={line width=.55pt},
 old/.style={bond,blue!70!cyan},
 new/.style={bond,green!65!black},
 cond/.style={paperred,line width=.65pt,densely dashed},
 sep/.style={black!70,line width=.45pt}]

\newcommand{\noedge}[2]{\draw[cond] (#1)--node[midway,fill=white,inner sep=.4pt,text=paperred,font=\bfseries] {$\times$}(#2);}

\begin{scope}[shift={(0,3.8)}]
\draw[sep] (2.00,-.15)--(2.00,2.05);
\draw[sep] (4.00,-.15)--(4.00,2.05);
\node[anchor=north west,font=\small] at (.02,2.02) {$L$};
\node[anchor=north west,font=\small] at (2.05,2.02) {$K$};
\node[anchor=north west,font=\small] at (4.05,2.02) {$R$};

\node[atom] (aL3) at (.95,1.55) {H};
\node[atom] (aL2) at (.35,1.08) {O};
\node[atom] (aL1) at (.35,.48) {C};
\node[atom] (aL0) at (.95,.18) {C};
\node[atom] (aL5) at (1.55,.48) {C};
\node[atom] (aL4) at (1.55,1.08) {O};
\draw[old] (aL2)--(aL3);
\draw[old] (aL1)--(aL2);
\draw[old,double,double distance=.9pt] (aL0)--(aL1);
\draw[old,double,double distance=.9pt] (aL4)--(aL5);
\noedge{aL0}{aL5}

\node[atom] at (2.95,1.55) {H};
\node[atom] at (2.35,1.08) {O};
\node[atom] at (2.35,.48) {C};
\node[atom] at (2.95,.18) {C};
\node[atom] at (3.55,.48) {C};
\node[atom] at (3.55,1.08) {O};

\node[atom] (aR3) at (4.95,1.55) {H};
\node[atom] (aR2) at (4.35,1.08) {O};
\node[atom] (aR1) at (4.35,.48) {C};
\node[atom] (aR0) at (4.95,.18) {C};
\node[atom] (aR5) at (5.55,.48) {C};
\node[atom] (aR4) at (5.55,1.08) {O};
\draw[new,double,double distance=.9pt] (aR1)--(aR2);
\draw[new] (aR0)--(aR1);
\draw[new] (aR0)--(aR5);
\draw[new] (aR4)--(aR5);
\draw[new] (aR3)--(aR4);
\node[font=\bfseries] at (3,-.48) {(a)};
\end{scope}

\begin{scope}[shift={(6.55,3.8)}]
\draw[sep] (2.00,-.15)--(2.00,2.05);
\draw[sep] (4.00,-.15)--(4.00,2.05);
\node[anchor=north west,font=\small] at (.02,2.02) {$L$};
\node[anchor=north west,font=\small] at (2.05,2.02) {$K$};
\node[anchor=north west,font=\small] at (4.05,2.02) {$R$};

\foreach \p/\x in {bL/0,bK/2,bR/4}{
  \node[atom] (\p3) at (\x+.95,1.55) {H};
  \node[atom] (\p2) at (\x+.35,1.08) {O};
  \node[atom] (\p1) at (\x+.35,.48) {C};
  \node[atom] (\p0) at (\x+.95,.18) {C};
  \node[atom] (\p5) at (\x+1.55,.48) {C};
  \node[atom] (\p4) at (\x+1.55,1.08) {O};
  \draw[bond] (\p2)--(\p3);
  \draw[bond] (\p1)--(\p2);
  \draw[bond,double,double distance=.9pt] (\p0)--(\p1);
  \draw[bond,double,double distance=.9pt] (\p4)--(\p5);
}
\noedge{bL0}{bL5}
\node[font=\bfseries] at (3,-.48) {(b)};
\end{scope}

\begin{scope}[shift={(0,0)}]
\draw[sep] (2.00,-.15)--(2.00,2.94);
\draw[sep] (4.00,-.15)--(4.00,2.94);
\node[anchor=north west,font=\small] at (.02,2.94) {$L$};
\node[anchor=north west,font=\small] at (2.05,2.94) {$K$};
\node[anchor=north west,font=\small] at (4.05,2.94) {$R$};

\foreach \p/\x in {cL/0,cK/2,cR/4}{
  \node[atom] (\p0) at (\x+.25,2.30) {C};
  \node[atom] (\p1) at (\x+.72,2.00) {C};
  \node[atom] (\p2) at (\x+1.25,2.30) {O};
  \node[atom] (\p3) at (\x+1.72,2.00) {H};
  \draw[bond,double,double distance=.9pt] (\p0)--(\p1);
  \draw[bond] (\p1)--(\p2)--(\p3);
}

\foreach \p/\x in {cL/0,cK/2,cR/4}{
  \node[atom] (\p6) at (\x+.72,1.32) {C};
  \node[atom] (\p5) at (\x+.25,1.02) {C};
  \node[atom] (\p7) at (\x+1.25,1.02) {C};
  \node[atom] (\p4) at (\x+.25,.35) {O};
  \node[atom] (\p8) at (\x+1.25,.35) {O};
  \node[atom] (\p9) at (\x+.72,.04) {H};
}

\draw[old,double,double distance=.9pt] (cL4)--(cL5);
\draw[old,double,double distance=.9pt] (cL6)--(cL7);
\draw[old] (cL7)--(cL8)--(cL9);

\draw[new] (cR4)--(cR5);
\draw[new] (cR6)--(cR5);
\draw[new] (cR6)--(cR7);
\draw[new,double,double distance=.9pt] (cR7)--(cR8);
\draw[new] (cR9)--(cR4);

\noedge{cL0}{cL5}
\node[font=\bfseries] at (3,-.48) {(c) $-1$};
\end{scope}

\begin{scope}[shift={(6.55,0)}]
\draw[sep] (2.00,-.15)--(2.00,2.94);
\draw[sep] (4.00,-.15)--(4.00,2.94);
\node[anchor=north west,font=\small] at (.02,2.94) {$L$};
\node[anchor=north west,font=\small] at (2.05,2.94) {$K$};
\node[anchor=north west,font=\small] at (4.05,2.94) {$R$};

\foreach \p/\x in {dL/0,dK/2,dR/4}{
  \node[atom] (\p6) at (\x+.25,2.30) {C};
  \node[atom] (\p7) at (\x+.72,2.00) {C};
  \node[atom] (\p8) at (\x+1.25,2.30) {O};
  \node[atom] (\p9) at (\x+1.72,2.00) {H};
  \draw[bond,double,double distance=.9pt] (\p6)--(\p7);
  \draw[bond] (\p7)--(\p8)--(\p9);
}

\foreach \p/\x in {dL/0,dK/2,dR/4}{
  \node[atom] (\p0) at (\x+.72,1.32) {C};
  \node[atom] (\p1) at (\x+1.25,1.02) {C};
  \node[atom] (\p5) at (\x+.25,1.02) {C};
  \node[atom] (\p2) at (\x+1.25,.35) {O};
  \node[atom] (\p4) at (\x+.25,.35) {O};
  \node[atom] (\p3) at (\x+.72,.04) {H};
}

\draw[old,double,double distance=.9pt] (dL0)--(dL1);
\draw[old] (dL1)--(dL2)--(dL3);
\draw[old,double,double distance=.9pt] (dL4)--(dL5);

\draw[new] (dR0)--(dR1);
\draw[new,double,double distance=.9pt] (dR1)--(dR2);
\draw[new] (dR0)--(dR5);
\draw[new] (dR5)--(dR4)--(dR3);

\draw[cond,rounded corners=1.5pt] (dL6.south)--(.25,1.67)
  --node[midway,fill=white,inner sep=.4pt,text=paperred,font=\bfseries] {$\times$}
  (1.25,1.67)--(dL1.north);
\node[font=\bfseries] at (3,-.48) {(d) $+1$};
\end{scope}

\end{tikzpicture}

\caption{Conditional composition for the aldol addition.
(a) Reaction rule $L\leftarrow K\to R$; (b) identity query with \NoEdge{};
(c,d) selected $r_{180}$ composites from rule-after-query and
query-after-rule composition, respectively, with coefficients $-1,+1$.
Corresponding atoms occupy the same positions.
The spans are isomorphic, but transported \NoEdge{} constraints differ:
the terms cancel only without the query's constraint.
Red crosses mark absence constraints; edge-creation admissibility is implicit.}
\label{fig:aldol-composition}
\end{figure}
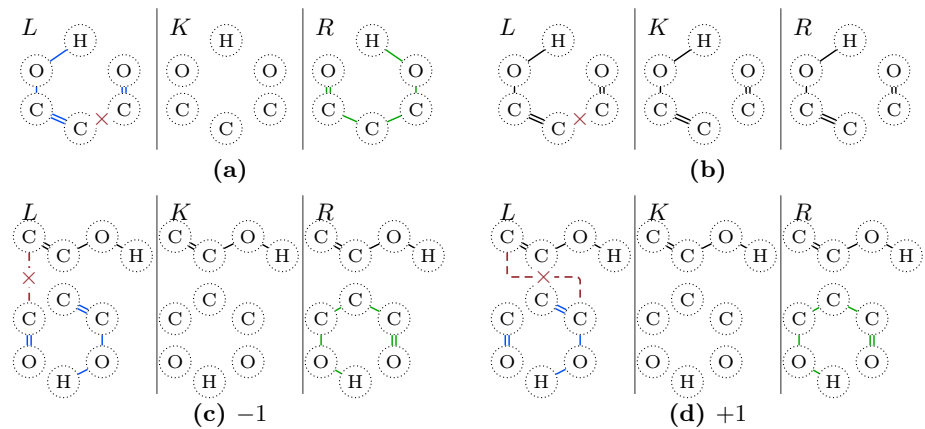

We revisit the aldol addition studied by Behr et
al.~\cite[Section 8.3]{BehrKrivineAndersenMerkle2021}. The observable counts
matches of a molecular pattern in which two designated carbon atoms must
not be bonded. We represent it as an identity rule with a \NoEdge{}
constraint between these atoms. The original study analyses the effect of
this condition manually; our implementation computes the conditional
compositions automatically. Figure~\ref{fig:aldol-composition} shows the
reaction rule, the identity rule for the observable and two selected
composites. \examplenotebook{01_aldol}{Aldol notebook}

We compose the reaction and identity rule in both orders, with and without
the identity rule's \NoEdge{} constraint. The reaction rule's own \NoEdge{}
constraint and the valence filter (atom degree weighted by bond order)
remain unchanged. Without the identity rule's constraint, we obtain
13 contributions when the reaction precedes the identity rule and 18 in
the reverse order. Adding the constraint
excludes three contributions in the first order, giving 10 and 18.

The constraint also changes which contributions cancel. Two composites
can have isomorphic underlying spans but different transported \NoEdge{}
constraints. They must then remain distinct when collecting the positive
and negative contributions. Consequently, although fewer compositions are
admissible, the number of nonzero signed terms in the commutator
(Equation~\eqref{eq:count-change}) increases
from 13 to 18.

Panels (c,d) show this effect for the published $r_{180}$ overlap. Without
the identity rule's \NoEdge{} constraint, the two selected terms cancel.
With it, the transported constraints refer to different atom pairs,
leaving two terms with coefficients $-1,+1$. Direct injective matching on
a 16-atom host graph gives input-pattern match counts $(0,1)$ with these
constraints, compared with $(1,1)$ when the identity rule's constraint is
removed. Cancelling solely by the underlying spans would therefore discard
a nonzero contribution.

For a separate condition-free control, we remove both \NoEdge{} constraints.
M\O D and MolFoundry give the same compositions, including one contribution
absent from the published table. The comparison above instead retains the
reaction rule's constraint. Using an explicit absence-edge
encoding, M\O D also gives the same conditional compositions after translation
to \NoEdge{} constraints. The
\href{\SourceRoot/examples/tcs/README.md}{accompanying material}
provides the rules, overlap maps, host graph and per-rule comparisons.

\subsection{Changing the Query with the Backend Held Fixed}
\label{sec:query-reuse}

We change only the query in the ungated A/B path model of
Section~\ref{sec:path-backend}. The four feed/append rules and backend
stay fixed while we select $AB$, $ABA$, $ABBA$, $AAAA$ and $ABABA$.
Each run allows at most 64 discovered patterns,
64 worklist expansions and 24 functions in the expectation basis.
Table~\ref{tab:query-reuse} lists the discovered end patterns and events.
Repeated-letter motifs require identities between overlapping suffix
conditions: a path ending in $AAA$ also ends in $AA$ and $A$, so
$S_{AAA}=1$ implies $S_{AA}=S_A=1$. The same counting procedures resolve
these dependencies when constructing the event classes.
\examplenotebook{08_query_variation}{Query-variation notebook}

\begin{table}[!tb]
\caption{Five sole queries, unchanged path rules and backend. An end word
$w$ denotes the discovered pattern $S_w$. Coordinates also include the selected
motif and two free-pool counts. Time is the median of three fresh serial
processes for support discovery and event extraction, excluding validation
and linear closure. Terms are raw composites before removing impossible
patterns and cancelling signed contributions. All event constructions complete; every linear search
reaches its 24-function budget.}
\label{tab:query-reuse}
\centering\small
\setlength{\tabcolsep}{4pt}
\begin{tabular}{@{}llrrr@{}}
\toprule
Query & Discovered end words & Coordinates / events & Terms & Time (s)\\
\midrule
$AB$ & $A$ & 4 / 6 & 118 & 0.074\\
$ABA$ & $A,AB$ & 5 / 8 & 334 & 0.128\\
$ABBA$ & $A,AB,ABB$ & 6 / 10 & 775 & 0.236\\
$AAAA$ & $A,AA,AAA$ & 6 / 10 & 775 & 0.230\\
$ABABA$ & $A,AB,ABA,ABAB$ & 7 / 12 & 1,557 & 0.445\\
\bottomrule
\end{tabular}
\end{table}
 
Direct enumeration of bounded words and token pools agrees with the
generated coordinate images and joint jump rates for every query.

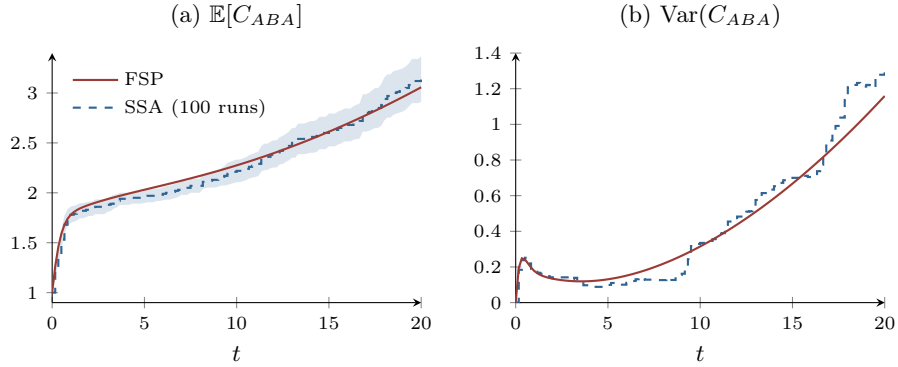
\begin{figure}[!tb]
\centering
\begin{tikzpicture}
\definecolor{abared}{HTML}{9C3B35}
\definecolor{abablue}{HTML}{2F6595}
\begin{groupplot}[
 group style={group size=2 by 1,horizontal sep=1.25cm},
 width=.40\textwidth,height=3.3cm,scale only axis,
 xmin=0,xmax=20,xtick={0,5,10,15,20},
 xlabel={$t$},axis lines=left,
 tick label style={font=\scriptsize},label style={font=\small},
 title style={font=\small},
 legend style={font=\scriptsize,draw=none,fill=none,
               at={(.03,.97)},anchor=north west},
 legend cell align=left]
\nextgroupplot[title={(a) $\mathbb E[C_{ABA}]$},ymin=.9,ymax=3.4,
 ytick={1,1.5,2,2.5,3}]
\addlegendimage{abared,thick}
\addlegendentry{FSP}
\addlegendimage{abablue,dashed,thick}
\addlegendentry{SSA (100 runs)}
\addplot[name path=abalower,draw=none,forget plot]
 table[x=t,y=ssa_mean_lower,col sep=comma]{aba-moments.csv};
\addplot[name path=abaupper,draw=none,forget plot]
 table[x=t,y=ssa_mean_upper,col sep=comma]{aba-moments.csv};
\addplot[abablue!17,draw=none,forget plot] fill between[of=abalower and abaupper];
\addplot[abablue,dashed,thick,const plot,forget plot]
 table[x=t,y=ssa_mean,col sep=comma]{aba-moments.csv};
\addplot[abared,thick,forget plot]
 table[x=t,y=mean,col sep=comma]{aba-moments.csv};
\nextgroupplot[title={(b) $\operatorname{Var}(C_{ABA})$},ymin=0,ymax=1.4,
 ytick={0,.2,.4,.6,.8,1,1.2,1.4}]
\addplot[abablue,dashed,thick,const plot]
 table[x=t,y=ssa_variance,col sep=comma]{aba-moments.csv};
\addplot[abared,thick]
 table[x=t,y=variance,col sep=comma]{aba-moments.csv};
\end{groupplot}
\end{tikzpicture}
 \caption{Mean and variance from the generated gated $ABA$ event model:
red curves are computed from the finite ODE approximation using FSP;
blue curves show estimates from 100 SSA runs. Shading shows
$\pm1.96$ Monte Carlo standard errors of the mean.
Initially the path is $ABAB$, $(M_A,M_B)=(2,1)$ and the gate is open.
Rates are $\beta_A=2$, $\beta_B=1.5$, $k_A=0.9$, $k_B=0.01$,
$a=0.6$ (close) and $b=0.4$ (open).}
\label{fig:aba-moments}
\end{figure}

For numerical illustration, we use the gated $ABA$ event model.
Figure~\ref{fig:aba-moments} compares finite-state projection
(FSP)~\cite{MunskyKhammash2006FSP} with the stochastic simulation algorithm
(SSA)~\cite{Gillespie1977SSA}. These standard numerical methods are
downstream of symbolic closure analysis. FSP restricts count ranges and
solves finite ODEs for probabilities and moments; SSA estimates the moments
from 100 independent trajectories.
The notebook defaults to the
ungated running example; setting \texttt{GATE=True} and \texttt{T=20}
reproduces this configuration.
\examplelinks{rop-family}{02_aba}{ABA}

\subsection{Recovering a Calculation on Unbounded Trees}
\label{sec:graph-examples}
\label{sec:remy}

We next analyse the planar rooted binary-tree model
\cite[Section 5]{Behr2021StochasticCombinatorics}. Its six rules subdivide
a selected edge, attach a leaf and choose the old subtree's left/right
placement. Edge labels $I,L,R$ distinguish the root attachment and left
and right children. The original uses sesqui-pushout (SqPO) rewriting; we represent
the same insertions by DPO rules on this tree family, with a common
per-match rate $\kappa$. The correspondence is explained in
\appref{app:remy}{B.2}.
\examplelinks{remy-trees}{04_remy}{R\'emy}

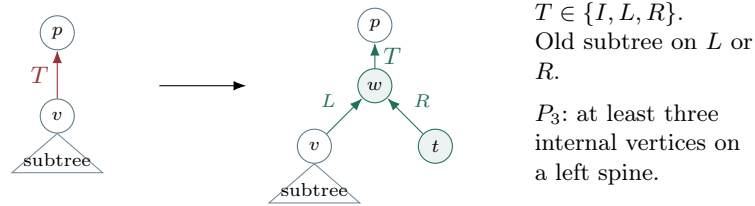
\begin{figure}[!tb]
\centering
\begin{tikzpicture}[x=1cm,y=1cm,>=Latex,font=\small,
 v/.style={circle,draw=paperblue!75,fill=white,minimum size=4.5mm,inner sep=0pt,font=\scriptsize},
 nv/.style={v,draw=papergreen,fill=papergreen!8}]
\begin{scope}[yshift=4.1cm]
\node[anchor=west,font=\small\bfseries] at (0,-.83) {R\'emy insertion: one of six typed rules};
\node[v] (oldparent) at (1.8,-1.65) {$p$};
\node[v] (oldroot) at (1.8,-2.75) {$v$};
\draw[->,paperred] (oldroot)--node[left]{$T$}(oldparent);
\draw[paperblue!65] (oldroot.south)--(1.2,-3.5)--(2.4,-3.5)--cycle;
\node[font=\scriptsize] at (1.8,-3.32) {subtree};
\draw[->] (3.15,-2.35)--(4.3,-2.35);
\node[v] (parent) at (6,-1.55) {$p$};
\node[nv] (newroot) at (6,-2.35) {$w$};
\node[v] (root) at (5.2,-3.15) {$v$};
\node[nv] (leaf) at (6.8,-3.15) {$t$};
\draw[->,papergreen] (newroot)--node[right]{$T$}(parent);
\draw[->,papergreen] (root)--node[above left,font=\scriptsize]{$L$}(newroot);
\draw[->,papergreen] (leaf)--node[above right,font=\scriptsize]{$R$}(newroot);
\draw[paperblue!65] (root.south)--(4.6,-3.9)--(5.8,-3.9)--cycle;
\node[font=\scriptsize] at (5.2,-3.72) {subtree};
\node[anchor=west,align=left,text width=3cm,font=\footnotesize] at (8,-2.45)
 {$T\in\{I,L,R\}$.\\Old subtree on $L$ or $R$.\\[5pt]
 $P_3$: at least three\\internal vertices on\\a left spine.};
\end{scope}
\end{tikzpicture}
\caption{Tree insertion~\cite[Equation (45)]{Behr2021StochasticCombinatorics}:
a new vertex $w$ subdivides $v\to p$ and gains a leaf $t$.
Green marks additions; triangles denote the preserved subtree.
Three edge types $T\in\{I,L,R\}$ and two subtree placements give six rules.}
\label{fig:remy}
\end{figure}

Let $P_j$ count rooted subtrees with at least $j$ internal
vertices on their left spine---the path obtained by repeatedly taking the
left child~\cite[Equations (45) and (54)]{Behr2021StochasticCombinatorics}.
Figure~\ref{fig:remy} shows one insertion rule. The only selected count
is $P_3$.

Composition and exact linear-dependence tests discover $P_2,P_1$ and the
edge count $E$. Algorithm~\ref{alg:linear} sums the six contributions in their
shared rate channel before testing the span. Writing
$\mu_E=\mathbb E[E]$ and $\mu_j=\mathbb E[P_j]$, it yields
\[
 \begin{aligned}
 \dot\mu_E&=4\kappa\mu_E,& \dot\mu_1&=2\kappa\mu_E,\\
 \dot\mu_2&=\kappa(3\mu_1-2\mu_2),&
 \dot\mu_3&=4\kappa(\mu_2-\mu_3).
 \end{aligned}
\]
The tree backend also computes four events by counting insertion contexts,
using an identity valid for all finite trees in this family.
The generated events also recover the published
Equation~(57g)~\cite{Behr2021StochasticCombinatorics}, using the conversion
given in \appref{app:remy}{B.2}.\footnote{Applying the stated change of
variables gives monomial factors different from those printed in
Equation~(58), apparently due to a typographical slip. The
\href{\SourceRoot/docs/remy-trees.md}{accompanying material} records the
comparison.}
Independent checks on concrete trees confirm the generated jumps and their
multiplicities. \appref{app:remy}{B.2} gives the context-counting
argument and the validation details.

\subsection{Eight Sites: Closure from Structural Invariants}
\label{sec:comparison}
\label{sec:eight-sites}

\begin{figure}[!t]
\centering
\begin{tikzpicture}[x=1cm,y=1cm,>=Latex,font=\small,
 v/.style={circle,draw=paperblue!75,fill=white,minimum size=4.5mm,inner sep=0pt,font=\scriptsize}]
\node[anchor=west] at (0,1.45) {(a)};
\foreach \i in {0,...,7}{
 \node[v] (c\i) at ($(1.05,.1)+(90-45*\i:.9)$) {$C$};
}
\foreach \i [evaluate=\i as \j using {int(mod(\i+1,8))}] in {0,...,7}
 \draw (c\i)--(c\j);
\foreach \i in {0,2,5}
 \node[v,fill=papergreen!55] at (c\i.center) {$C$};
\node at (1.05,-1.35) {$P_0=3$};

\node[anchor=west] at (3,1.45) {(b) $r_{\rm on}$};
\foreach \x/\s in {4.1/u,8.25/b}{
 \node[v] (a\s) at (\x,.9) {$C$};
 \node[v] (p\s) at (\x-.45,.2) {$P$};
 \node[v] (l\s) at (\x+.45,.2) {$L$};
 \draw (a\s)--(p\s);
 \draw (a\s)--(l\s);
}
\path (au)--node[left,font=\tiny]{cp}(pu);
\path (au)--node[right,font=\tiny]{cl}(lu);
\draw[paperred,dashed] (pu)--node[below=2.5mm,font=\scriptsize]{\NoEdge{}}(lu);
\draw[papergreen,thick] (pb)--node[below=2.5mm,font=\scriptsize]{\texttt{bound}}(lb);
\draw[->] (5.5,.55)--(6.9,.55);

\node[anchor=west] at (3,-.6) {(c) $r_{\rm coop}$};
\foreach \x/\s in {3.5/t,5/n,7.8/T,9.3/N}{
 \node[v] (a\s) at (\x,-1.1) {$C$};
 \node[v] (p\s) at (\x-.4,-1.8) {$P$};
 \node[v] (l\s) at (\x+.4,-1.8) {$L$};
 \draw (a\s)--(p\s);
 \draw (a\s)--(l\s);
}
\draw (at)--node[above,font=\scriptsize]{\texttt{next}}(an);
\draw (aT)--node[above,font=\scriptsize]{\texttt{next}}(aN);
\draw[paperred,dashed] (pt)--node[below=2.5mm,font=\scriptsize]{\NoEdge{}}(lt);
\draw (pn)--(ln);
\draw (pN)--(lN);
\draw[papergreen,thick] (pT)--(lT);
\draw[->] (5.8,-1.45)--(6.9,-1.45);
\end{tikzpicture}
\caption{Eight-site model. (a) An eight-cycle is one permitted topology;
two four-cycles and a three-cycle plus a five-cycle are also allowed.
Filled sites are occupied; \texttt{next} edges join neighbouring sites. Each site has a $C,P,L$ triple, expanded in
(b)--(c), with fixed \texttt{cp} and \texttt{cl} edges. The optional
\texttt{bound} edge joins $P$ and $L$; these labels are abstract types.
(b) Ordinary binding creates that edge (green) only when
\NoEdge$(P,L)$ holds (red dashed line). (c) Cooperative binding additionally
matches an occupied neighbour, whose triangle is unchanged. Each occupied
neighbour gives a separate match. Unbinding deletes the \texttt{bound} edge.
The query $P_0$ counts complete $C,P,L$ triangles, hence occupied sites.}
\label{fig:ring8}
\end{figure}
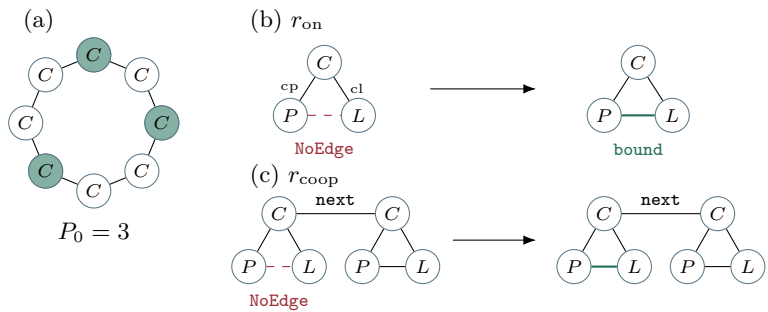
 This case tests whether \NoEdge{}-guarded binding yields closed equations
without enumerating occupancy states. The input specifies eight owned sites,
each with exactly two distinct \texttt{next} neighbours, three binding/unbinding
rules and the occupied-site query (Figure~\ref{fig:ring8}). The simple
undirected site graph permits three topologies: an eight-cycle, two four-cycles,
or a three-cycle and a five-cycle. The backend completes
connected path patterns to their possible cycle components and expands
missing-edge tests into polynomial count identities
(\appref{app:eight-slots}{B.3}). Incidence bounds implied by $\Omega$ prune
impossible overlaps before composition. From the single query, composition discovers
a closed 48-coordinate expectation system; a subsequent rank check finds
one redundant coordinate, giving function-space dimension 47.
Although exhaustive enumeration is infeasible for unbounded state spaces,
here we use it after synthesis to independently verify the generated dynamics
on all $2^8=256$ free/occupied assignments per topology ($768$ total). All three topologies use the same ODE, with initial
values obtained by counting patterns in the starting graph.
\examplelinks{eight-slots}{05_eight_slots}{Eight-site}

\subsection{Comparison with FRAGGER}
\label{sec:fragger}

\paragraph{The published DNA walker.}
For comparison with FRAGGER, we use the two-legged DNA walker~\cite[Section 1.1]{DanosHeindelHonoratoZimmerStucki2020}.
We consider their single-walker case on a circular DNA scaffold, with the
two legs at most one DNA segment apart, so that the walker is either
compressed or extended. Four graph rules describe forward and backward
leg movements. Starting from only the extended-configuration count,
MolFoundry composition introduces the compressed count and closes a
two-dimensional linear expectation system. The retained FRAGGER output contains the same two
equations from its original directed multigraph representation, whereas
MolFoundry represents directed edges by auxiliary vertices and uses rule
composition. All eight
contributions of the four rules to the two observables agree, with zero
discrepancies. The notebook recomputes this discovery through
Algorithm~\ref{alg:linear} with a path-extension counting identity.
The compressed pattern has two injective matches that exchange
its two leg edges; after accounting for this factor of two
(\appref{app:dna-walker}{B.4}), both approaches also give the same stationary
mean velocity as derived in~\cite{DanosHeindelHonoratoZimmerStucki2020}.
\examplenotebook{06_dna_walker}{DNA-walker notebook}

\paragraph{Shared composition workload.}
For an $n$-vertex alternating A/B query, median times in seconds for twelve
ordered rule compositions are:
\par\smallskip
\noindent\begin{minipage}{\linewidth}
\centering\small
\setlength{\tabcolsep}{4pt}
\begin{tabular}{@{}lrrrrrr@{}}
\toprule
$n$ & 4 & 12 & 20 & 24 & 26 & 28\\
\midrule
MolFoundry 0.1.9 & 0.0102 & 0.0364 & 0.0987 & 0.15 & 0.18 & 0.213\\
FRAGGER & 0.00804 & 0.0561 & 4.32 & 74.3 & 314 & 1300\\
\bottomrule
\end{tabular}
\end{minipage}
\par\smallskip
 All twelve per-order raw counts agree through $n=28$. Protocol: \appref{app:composition-benchmark}{C}.
\examplenotebook{09_composition_comparison}{Benchmark notebook}

\subsection{From A/B Paths to Atom-Resolved Polymer Graphs}
\label{sec:molecular-validation}
\label{sec:molecular-applications}

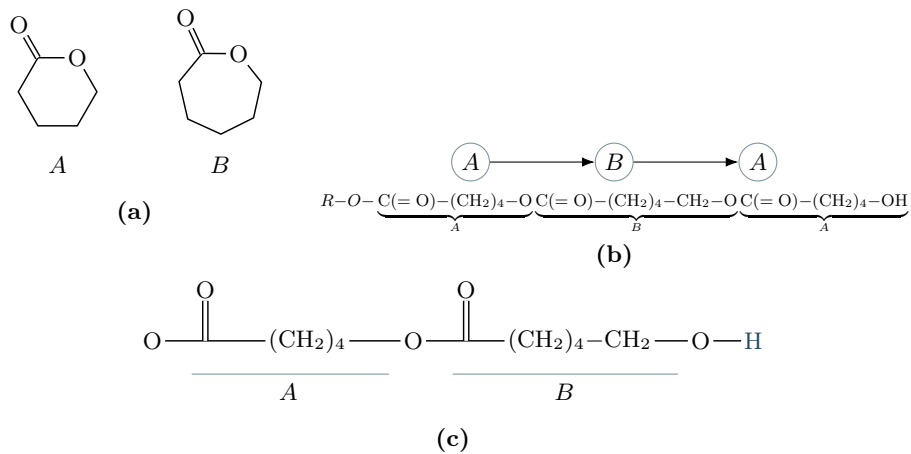
\begin{figure}[!tb]
\centering
\begin{minipage}[t]{0.31\linewidth}
\centering
\begin{tikzpicture}[x=1cm,y=1cm,font=\small,
 atom/.style={fill=white,inner sep=1pt},
 bond/.style={line width=.55pt}]
\foreach \i/\ang in {0/120,1/60,2/0,3/-60,4/-120,5/180}
 \coordinate (a\i) at ($(0,.25)+(\ang:.52)$);
\node[atom] (ao) at (a1) {O};
\draw[bond] (a0)--(ao)--(a2)--(a3)--(a4)--(a5)--(a0);
\node[atom] (aoxo) at ($(0,.25)+(120:1.02)$) {O};
\draw[double,double distance=1.25pt,bond] (a0)--(aoxo);
\node[font=\footnotesize] at (0,-.72) {$A$};

\foreach \i/\ang in {0/115,1/64,2/13,3/-38,4/-89,5/-140,6/166}
 \coordinate (b\i) at ($(2.15,.25)+(\ang:.56)$);
\node[atom] (bo) at (b1) {O};
\draw[bond] (b0)--(bo)--(b2)--(b3)--(b4)--(b5)--(b6)--(b0);
\node[atom] (boxo) at ($(2.15,.25)+(115:1.08)$) {O};
\draw[double,double distance=1.25pt,bond] (b0)--(boxo);
\node[font=\footnotesize] at (2.15,-.72) {$B$};
\end{tikzpicture}
\par\smallskip\textbf{(a)}
\end{minipage}\hfill
\begin{minipage}[t]{0.65\linewidth}
\centering
\begin{tikzpicture}[x=1cm,y=1cm,font=\small,>=Latex]
\node[circle,draw=paperblue!60,minimum size=5mm,inner sep=0pt] (p1) at (.75,.85) {$A$};
\node[circle,draw=paperblue!60,minimum size=5mm,inner sep=0pt] (p2) at (2.65,.85) {$B$};
\node[circle,draw=paperblue!60,minimum size=5mm,inner sep=0pt] (p3) at (4.55,.85) {$A$};
\draw[->] (p1)--(p2);
\draw[->] (p2)--(p3);
\end{tikzpicture}

\vspace{-.5ex}
\resizebox{.98\linewidth}{!}{$
R{-}O{-}
\underbrace{\mathrm{C(=O){-}(CH_2)_4{-}O}}_{A}
\underbrace{\mathrm{C(=O){-}(CH_2)_4{-}CH_2{-}O}}_{B}
\underbrace{\mathrm{C(=O){-}(CH_2)_4{-}OH}}_{A}
$}
\par\smallskip\textbf{(b)}
\end{minipage}

\smallskip

\begin{minipage}[t]{\linewidth}
\centering
\begin{tikzpicture}[x=1cm,y=1cm,font=\small,
 atom/.style={fill=white,inner sep=1pt},bond/.style={line width=.55pt}]
\node[atom] (iO) at (.4,.2) {O};
\coordinate (cA) at (1.1,.2);
\node[atom] (oA) at (1.1,.88) {O};
\node[atom] (a) at (2.45,.2) {$(\mathrm{CH}_2)_{4}$};
\node[atom] (mid) at (3.85,.2) {O};
\coordinate (cB) at (4.55,.2);
\node[atom] (oB) at (4.55,.88) {O};
\node[atom] (b) at (6.05,.2) {$(\mathrm{CH}_2)_{4}{-}\mathrm{CH}_2$};
\node[atom] (end) at (7.65,.2) {O};
\node[atom,text=paperblue] (h) at (8.35,.2) {H};
\draw[bond] (iO)--(cA)--(a)--(mid)--(cB)--(b)--(end)--(h);
\draw[double,double distance=1.4pt,bond] (cA)--(oA);
\draw[double,double distance=1.4pt,bond] (cB)--(oB);
\draw[paperblue!60] (.92,-.24)--(3.53,-.24);
\draw[paperblue!60] (4.37,-.24)--(7.35,-.24);
\node[font=\footnotesize] at (2.2,-.48) {$A$};
\node[font=\footnotesize] at (5.85,-.48) {$B$};
\end{tikzpicture}
 \par\smallskip\textbf{(c)}
\end{minipage}

\caption{From repeat sequences to molecular patterns.
(a) Cyclic monomers:
$A=\delta$-valerolactone and $B=\varepsilon$-caprolactone.
(b) Abstract $A\!\to\!B\!\to\!A$ and condensed molecular $ABA$, with
$B$'s extra methylene group shown explicitly.
(c) Discovered end pattern $S_{AB}$, anchored by its terminal O--H group.
Hosts use explicit C, O and H vertices; methylene groups are condensed
only for readability.}
\label{fig:molecules}
\label{fig:rop-net-rule}
\end{figure}

We now repeat the A/B path analysis directly on molecular graphs. Here
$A$ is $\delta$-valerolactone and $B$ is $\varepsilon$-caprolactone;
ring-opening polymerization (ROP) incorporates the atoms of a free lactone
into a benzyl-initiated, unbranched chain~\cite{NeilsonBielawski2013,WangHanZhang2020}.
Figure~\ref{fig:rop-net-rule} contrasts the abstract $ABA$ representation
with the atom-resolved chain and one molecular end pattern.
\examplenotebook{07_molecular_rop}{Molecular ROP notebook}

Here composition starts from C, O and H atoms and covalent bonds;
the path model starts from repeat labels and sequence order.
Names such as $S_{AB}$ are assigned after the resulting molecular patterns
have been recognised. Under the graph-count correspondence, activities,
generator images and joint jump/rate distributions agree exactly with
the A/B model.

Repeated atom labels and larger patterns admit more overlaps. Local bond
restrictions exclude overlap candidates incompatible with the supplied
valence and bond constraints~\cite[Definition~19 and Example~7]{BehrKrivineAndersenMerkle2021}.
The molecular backend uses these restrictions during enumeration and reduces
surviving compositions to counts of free lactones, incorporated repeats and
hydroxyl-ended chain patterns.

As a separate check for ungated $ABA$, MolFoundry enumerates rule matches
on 153 molecular hosts, applies each match, and counts the selected and
discovered observables. All 612 rate-channel joint jump distributions agree
with the composition-derived activities and jumps. A separate symbolic
comparison with the abstract path model gives the same coarse event dynamics.
The all-chain counting arguments are given in
\appref{app:molecular-rop}{B.5}.

\subsection{A Neighbour NAC on an Unbounded Molecular Graph Family}
\label{sec:chlorination}

The molecule is an alkylcyclopentadienyl anion with four
C--H sites eligible for chlorination and an alkyl tail of arbitrary finite
length. Each chlorination replaces one C--H bond by C--Cl, consumes one HCl
molecule and produces one H$_2$ molecule. A separate rule grows the
alkyl tail. These net rules define the benchmark and are not intended as an
experimentally validated elementary mechanism.
\examplelinks{chlorination/context-nac}{03_chlorination}{Chlorination}

For this example, we assume that a ring carbon can be chlorinated only if
no neighbouring ring carbon already carries chlorine. This is expressed by a negative application condition
(NAC): there must not exist a neighbouring carbon bonded to chlorine.
A \NoEdge{} test between atoms already matched by the rule cannot express
this condition. We compare the model with and without this NAC, leaving
everything else unchanged.

The only selected query $Z$ counts ring C--Cl bonds. With total chlorine
inventory $H_0=4$, conditional composition and support discovery introduce
three positive pattern counts $P_1,P_2,P_3$. These count embeddings of
specific ring-scaffold patterns requiring two, three and four C--H sites,
respectively---not arbitrary subsets of free sites. For $P_1$, the two free
sites must be adjacent, with one neighbouring the charged anchor; the
initially unchlorinated ring admits two such embeddings, one on either
side of the anchor. Symbolic source matches
are partitioned by their complete coordinate jumps, giving four event
classes on $(Z,P_1,P_2,P_3)$. Without the NAC, $Z$ alone gives one event
class of activity $(4-Z)^2$. Both finite event models close; a separate
linear expectation-closure search is not performed in this analysis.
The locality certificate establishes that
tail growth has zero jump on these coordinates for arbitrary finite tails.
Post-synthesis checks cover all 16 substitution patterns in three molecular
contexts: direct MolFoundry applications and independent site updates with
NetworkX counts agree on the joint jumps and rates.

\begin{figure}[!tb]
\centering
\begin{minipage}[c]{.30\linewidth}
\centering\small
\textbf{Chlorination blocked}\\[2mm]
\begin{tikzpicture}[font=\small]
\foreach \i in {0,...,4}
 \coordinate (c\i) at (90-72*\i:.75);
\node[fill=white,inner sep=1pt] (c0a) at (c0) {$\mathrm C^{-}$};
\node[fill=white,inner sep=1pt] (tail) at (0,1.43) {$R$};
\draw (c0a)--(tail);
\draw (c0a)--(c1)--(c2)--(c3)--(c4)--(c0a);
\foreach \i/\j in {0/1,1/2,2/3,3/4,4/0}
 \draw[densely dashed] ($(c\i)!.18!(0,0)$)--($(c\j)!.18!(0,0)$);
\foreach \i/\txt/\col in {1/Cl/papergreen,2/H/paperred,3/H/paperred,4/Cl/papergreen}{
 \node[fill=white,inner sep=1pt,text=\col] (e\i)
    at (90-72*\i:1.21) {\txt};
 \draw (c\i)--(e\i);
}
\draw[paperred,line width=1.1pt] (c1)--(c2);
\draw[paperred,line width=1.1pt] (c3)--(c4);
\end{tikzpicture}\\[1mm]
\footnotesize Each free site has a\\chlorinated neighbour.\\
Neither can react.
\end{minipage}\hfill
\begin{minipage}[c]{.65\linewidth}
\centering
\begin{tikzpicture}
\begin{groupplot}[
 group style={group size=1 by 2,vertical sep=.65cm},
 width=\linewidth,height=2.65cm,
 xmin=0,xmax=3,xtick={0,1,2,3},
 tick label style={font=\scriptsize},label style={font=\small},
 axis lines=left,grid=major,grid style={paperblue!10},
 legend style={draw=none,fill=none,font=\scriptsize,
               at={(.5,-.08)},anchor=north},
 legend columns=2,
 legend cell align=left]
\nextgroupplot[ymin=0,ymax=4.2,ytick={0,2,4},ylabel={$\mathbb E[Z]$},
 xticklabels={}]
\addplot[papergreen,thick] table[x=tau,y=mean_nac,col sep=comma]
 {chlorination-nac.csv};
\addlegendentry{with NAC}
\addplot[paperred,dashed,thick] table[x=tau,y=mean_ablated,col sep=comma]
 {chlorination-nac.csv};
\addlegendentry{without NAC}
\nextgroupplot[ymin=0,ymax=1.05,ytick={0,.5,1},
 ylabel={$\Pr(\text{blocked})$},xlabel={$\tau=\kappa t$}]
\addplot[papergreen,thick] table[x=tau,y=blocked_nac,col sep=comma]
 {chlorination-nac.csv};
\addplot[paperred,dashed,thick] table[x=tau,y=blocked_ablated,col sep=comma]
 {chlorination-nac.csv};
\end{groupplot}
\end{tikzpicture}
\end{minipage}
\caption{Chlorination with and without the neighbour NAC.
Left: two chlorinated sites block both free sites; dashed bonds are
aromatic and $R$ is the preserved tail.
Right: mean ring chlorine count and probability of blocked chlorination
with free sites remaining, from the event models' finite forward equations.
Tail growth may continue.}
\label{fig:chlor-summary}
\end{figure}
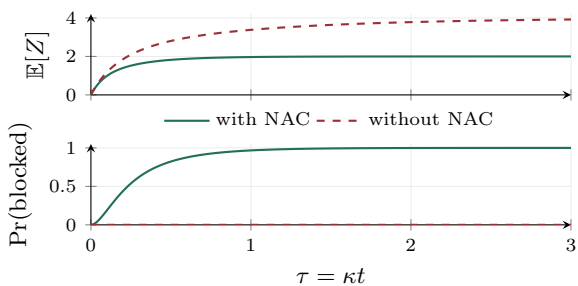
 
Figure~\ref{fig:chlor-summary} starts from an unchlorinated ring and four
HCl molecules, with $(Z,P_1,P_2,P_3)=(0,2,2,2)$. Initially all four sites
can react, each at rate $4\kappa$. For $\kappa>0$, the NAC arrests
chlorination at two ring chlorines with probability one; without it, all
four sites eventually react. The curves come from the finite forward
equations of the generated event models, explored only after symbolic
synthesis. They describe the selected counts, not a finite enumeration of
the full molecular state space. Tail growth may continue after chlorination
is blocked. The events and forward equations are given in
\chlorinationdetails{}.

\paragraph{Resources and limits.}
\label{sec:query-cost}
Larger queries can make composition expensive or exhaust closure budgets.
Query-variation timings do not isolate pruning costs or compare discovery
strategies; the FRAGGER benchmark times composition, not complete analysis.
Infinite-family counting identities rely on appendix arguments, not bounded checks alone.
The \href{\MaterialsURL}{material guide} includes an unfinished growing-ring
search and a Brusselator--ROP model, distinguishing numerical approximations
from exact generated equations.
 \section{Conclusion}
\label{sec:discussion}

We introduced an automated framework for observable analysis with
MolFoundry. It combines conditional composition with nested application
conditions and \NoEdge{}, auxiliary-observable discovery, and searches for
linear expectation closure and polynomial-jump-closed event models.
Exactness relies on supplied graph-family counting identities and jump
arguments. The examples reproduce and extend published analyses and
demonstrate chemical applications without enumerating unbounded reaction
spaces. Future work includes automating the derivation of counting
identities and extending the analysis to further graph families.
We are also developing native support for directed multigraphs, avoiding
the current incidence encoding of directed and parallel edges.
 \paragraph{Acknowledgements.}
This work was supported in part by the Novo Nordisk Foundation through
the MATOMIC project (grant NNF21OC0066551).
\paragraph{Data availability.}
The artifact will be archived on \href{https://zenodo.org/}{Zenodo}
(DOI pending); sources and Docker-based reproduction instructions are
available on \href{\PaperGuideURL}{GitHub}.
Interactive examples and executable Jupyter notebooks are available on\\
\url{https://stochmech.algochem.techfak.de/}.
\FloatBarrier

\clearpage
\appendix
\renewcommand{\theHsection}{math.\Alph{section}}
\renewcommand{\theHequation}{\theHsection.\arabic{equation}}
\section{Conditional Counts and Moment Equations}
\label{app:analysis-details}

\subsection{Conditions and a Worked Count Change}
\label{app:conditions}

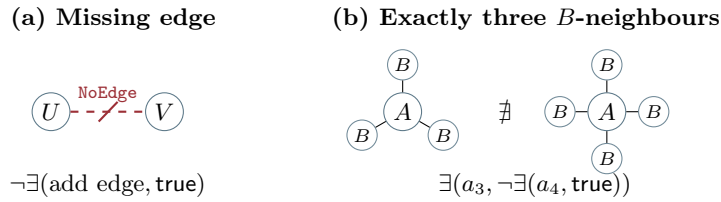
\begin{figure}[!tb]
\centering
\begin{tikzpicture}[x=1cm,y=1cm,>=Latex,font=\small,
 vertex/.style={circle,draw=paperblue!75,fill=white,minimum size=5mm,inner sep=0pt},
 neighbour/.style={circle,draw=paperblue!65,fill=white,minimum size=4mm,inner sep=0pt,font=\scriptsize}]
\node[anchor=west,font=\small\bfseries] at (0,.48) {(a) Missing edge};
\node[vertex] (cu) at (.65,-.75) {$U$};
\node[vertex] (cv) at (2.15,-.75) {$V$};
\draw[dashed,paperred,thick] (cu)--node[above,font=\scriptsize]{\NoEdge}(cv);
\draw[paperred,thick] (1.28,-.88)--(1.53,-.62);
\node[align=center,font=\footnotesize] at (1.4,-1.72)
 {$\neg\exists(\text{add edge},\mathsf{true})$};
\node[anchor=west,font=\small\bfseries] at (4.25,.48) {(b) Exactly three $B$-neighbours};
\node[vertex] (a3) at (5.3,-.75) {$A$};
\foreach \ang/\i in {90/1,210/2,330/3}{
 \node[neighbour] (b\i) at ($(a3)+(\ang:.62)$) {$B$};
 \draw (a3)--(b\i);
}
\node at (6.65,-.75) {$\nexists$};
\node[vertex] (a4) at (8,-.75) {$A$};
\foreach \ang/\i in {90/1,180/2,270/3,0/4}{
 \node[neighbour] (d\i) at ($(a4)+(\ang:.62)$) {$B$};
 \draw (a4)--(d\i);
}
\node[align=center,font=\footnotesize] at (7.05,-1.72)
 {$\exists(a_3,\neg\exists(a_4,\mathsf{true}))$};
\end{tikzpicture}
\caption{(a) \NoEdge{} forbids an edge between $U,V$.
(b) The anchored $A$ has three distinct $B$-neighbours and no fourth;
$a_3,a_4$ add three neighbours and then a fourth.
Dashed red lines mark absence tests. All matches are injective.}
\label{fig:conditional-rules}
\end{figure}

Nested conditions~\cite{HabelPennemann2009} on $P$ are built from
$\mathsf{true}_P$, negation, conjunction and
$\exists(a:P\hookrightarrow Q,c_Q)$. A match $m:P\hookrightarrow G$
satisfies the existential clause if an injective extension
$n:Q\hookrightarrow G$ exists with $n\circ a=m$ and $n\models c_Q$
\cite[Definition 2]{BehrKrivineAndersenMerkle2021}.
Figure~\ref{fig:conditional-rules} illustrates absence and exact degree:
$d(v)=k$ is $d(v)\geq k\land\neg(d(v)\geq k+1)$, with eligible labels
and bond types specified.

Let $u$ delete the \texttt{bound} edge between $U,V$, and let $O_X$ count
these pairs with no edge. Write $D$ for the empty-overlap composite and
$P_U,P_V$ for the composites identifying only one endpoint. The two orders give
\[
 O_X\star u=D+u+P_U+P_V,\qquad
 u\star O_X=D+u_\bot+P_U+P_V.
\]
In the full overlap, deletion establishes the missing-edge condition;
the reverse order gives $u_\bot$, requiring an edge both present and absent.
Cancellation and projection leave only the activity of $u$.
On the two-state family with one $U,V$ pair and only the optional
\texttt{bound} edge between them, this activity is $1-X$,
and $\partial_t\mathbb E[X]|_u=\kappa_u(1-\mathbb E[X])$.

 \FloatBarrier

\subsection{From Events to Moment Equations}
\label{app:emgf}

For a count vector $\phi$, the joint exponential moment-generating function
(EMGF) uses one variable in $\lambda$ per coordinate:
\begin{equation}
 \begin{aligned}
 M(t,\lambda)&=\mathbb E[e^{\lambda\cdot\phi(G_t)}],\qquad
 \partial_tM=\mathcal DM,\\
 \mathcal D&=\sum_e\kappa_{\gamma(e)}(e^{\lambda\cdot\delta_e}-1)
                              a_e(\partial_\lambda).
 \end{aligned}
 \label{eq:emgf}
\end{equation}
In $a_e(\partial_\lambda)$, each coordinate variable is replaced by the
partial derivative with respect to its corresponding $\lambda$ variable;
$\delta_e$ determines the exponential factor
\cite[Theorem 4.2]{BehrDanosGarnier2020}. The same moment calculus applies
to conditional pattern counts~\cite[Theorem 7]{BehrKrivineAndersenMerkle2021}.
Derivatives at $\lambda=0$ give joint moments of the counts, even when
no finite set of their equations closes.

The process interpretation requires nonexplosion, existence of the
relevant moments and interchange of generator with expectation. The EMGF
also requires convergence and the indicated differentiations. Otherwise,
Equation~\eqref{eq:emgf} is a formal power series when its moment identities
hold.
 \FloatBarrier
\section{Derivations for the Examples}
\label{app:examples}
\subsection{A/B Path and Missing-Edge Gate}
\label{app:path-equations}

On the rooted directed A/B-path family of
Section~\mainref{sec:workflow}, $q=C_{ABA}$ counts contiguous
$ABA$ occurrences, including overlaps; $M_A,M_B$ count free tokens and $S_w$ indicates an ending $w$.
Free tokens, persistent feed markers and the optional gate are disjoint
from the path. Thus a free-token factor and a pattern at its unique end
have product count $M_AS_w$ or $M_BS_w$.
The reducer accepts one connected path factor, free-token factors and
disjoint feed or gate factors. Repeated tokens of the same type use
falling-factorial counts; other overlap corrections remain unsupported.

A directed edge is encoded by an auxiliary vertex with distinct
\texttt{src}/\texttt{tgt} links to its endpoints. Each deleted terminal-edge
auxiliary vertex has exactly the two incidence edges included in the insertion match.
The DPO dangling condition is therefore automatic, and the projected
admissible-match count reduces to the full-input identity on this family.
The uniform-jump argument below applies to the checked feed, insertion and
gate templates, including their neighbour and \NoEdge{} conditions.

Use the discovered coordinate order
$\phi=(q,M_A,M_B,S_{AB},S_A)^{\mathsf T}$ and write $z=\phi(G)$.
The unique path end gives $S_A^2=S_A$, $S_{AB}^2=S_{AB}$ and
$S_AS_{AB}=0$. The main text gives the discovery and generator images
in Equation~(\mainref{eq:discovered-aba}).

Table~\ref{tab:path-events} uses this same coordinate order. \emph{Neither}
means that $S_A=S_{AB}=0$. Each row records the numerical jump $\delta_e$
and the full rate $\kappa_{\gamma(e)}a_e$ of Equation~(\mainref{eq:events}).
\begin{table}[!tb]
\caption{Eight events of the ungated $ABA$ model. The gated model adds
the coordinate $X$, gates the six append rates and adds opening and closing.}
\label{tab:path-events}
\centering\small
\setlength{\tabcolsep}{3pt}
\begin{tabular}{@{}lrrrrrl@{}}
\toprule
Rule / ending & $\Delta q$ & $\Delta M_A$ & $\Delta M_B$
 & $\Delta S_{AB}$ & $\Delta S_A$ & Full rate\\
\midrule
Feed $A$ &0&1&0&0&0&$\beta_A$\\
Feed $B$ &0&0&1&0&0&$\beta_B$\\
$r_A$ / neither &0&$-1$&0&0&1&$k_AM_A(1-S_A-S_{AB})$\\
$r_A$ / $A$ &0&$-1$&0&0&0&$k_AM_AS_A$\\
$r_A$ / $AB$ &1&$-1$&0&$-1$&1&$k_AM_AS_{AB}$\\
$r_B$ / neither &0&0&$-1$&0&0&$k_BM_B(1-S_A-S_{AB})$\\
$r_B$ / $A$ &0&0&$-1$&1&$-1$&$k_BM_BS_A$\\
$r_B$ / $AB$ &0&0&$-1$&$-1$&0&$k_BM_BS_{AB}$\\
\bottomrule
\end{tabular}
\end{table}
The three ending classes cover every finite path, including the empty one.
All free-token choices for a fixed insertion give the same jump. Dividing
the coordinate images by the rule activity therefore recovers the individual
increments wherever the rule is enabled.

For example, $ABAB$ with pools $(M_A,M_B)=(2,1)$ has coordinate values
$z=(1,2,1,1,0)^{\mathsf T}$. Inserting $A$ gives $ABABA$ and
$z'=(2,1,1,0,1)^{\mathsf T}$. Thus
\[
 \delta_e=z'-z=(1,-1,0,-1,1)^{\mathsf T},\qquad a_e(z)=2.
\]
The existing $ABA$ is preserved and another, overlapping occurrence is
created. The full rate is $2k_A$.

For the joint EMGF, write
$\theta_e=\lambda_q-\lambda_{M_A}-\lambda_{S_{AB}}+\lambda_{S_A}$.
The event contributes
\begin{equation}
 k_A(e^{\theta_e}-1)
 \partial_{\lambda_{M_A}}\partial_{\lambda_{S_{AB}}}M
 \label{eq:app-path-emgf}
\end{equation}
by combinatorial conversion~\cite[Theorem 4.2]{BehrDanosGarnier2020}.
The analogous terms for the other seven events complete the operator.
Its first moment involves $\mathbb E[M_AS_{AB}]$, explaining why
expectation closure requires a separate search.

With the missing-edge gate, $X=1$ denotes absence of the $U,V$ bond.
Both insertion rules require \NoEdge$(U,V)$. The discovered vector becomes
$\phi^+=(q,M_A,M_B,S_{AB},S_A,X)^{\mathsf T}$, with six coordinates.
All six insertion rates in Table~\ref{tab:path-events}
acquire a factor $X$, and their jump vectors acquire a zero final
coordinate. Feeds keep their rates and also acquire a zero final jump
coordinate. Closing and opening contribute
\[
 \delta_{\rm close}=(0,0,0,0,0,-1)^{\mathsf T},\quad aX,
 \qquad
 \delta_{\rm open}=(0,0,0,0,0,1)^{\mathsf T},\quad b(1-X),
\]
where $a,b$ are the closing and opening rate parameters, so the second
entry of each pair is the full rate. There are ten events
in total, and $\Gen X=b(1-X)-aX$. This gated model is used for the
numerical mean and variance curves in the main text. In Equation~\eqref{eq:app-path-emgf}, the insertion term gains
$\partial_{\lambda_X}$ while its exponential factor is unchanged.
The generated event tables and their derivations are available through the
\href{\SourceRoot/examples/aba/README.md}{ABA example guide}.
 \FloatBarrier
\subsection{R\'emy Trees}
\label{app:remy}

The rules, edge types and pattern observables come from
Behr~\cite[Section 5, Equations (45) and (54)]{Behr2021StochasticCombinatorics}.
Here $E$ counts edges and $P_j$ counts rooted subtrees with at least $j$
internal vertices on the path obtained by repeatedly taking the left child.
The analysis starts with only $P_3$.
Composition and exact linear dependence tests discover $P_2,P_1,E$.
All six insertions share one rate $\kappa$. Their generator contributions
are summed before the span test in Algorithm~\mainref{alg:linear}. This
four-function closure is for that channel; independent rates for the six
rules require a different closure search. The event construction below
uses its tree-context argument, without assuming uniform jumps for the
unrefined tree rules.

In the incidence encoding, a selected tree edge is an auxiliary vertex
with exactly two matched \texttt{src}/\texttt{tgt} links. Its deletion
satisfies the DPO dangling condition. Each insertion therefore has the
same matches and successor as Behr's SqPO rule, and its projected DPO
count reduces to the full-input identity on this tree family.

\begin{table}[!tb]
\caption{Generated R\'emy events. Full rates are $\kappa a_e$ for the
common per-match rate $\kappa$, in coordinate order $(E,P_1,P_2,P_3)$.}
\label{tab:remy-events}
\centering\small
\begin{tabular}{@{}ll@{}}
\toprule
Jump $\delta_e$ & Activity $a_e$\\
\midrule
$(2,1,0,0)$  & $2E-3P_1+2P_2-P_3$\\
$(2,1,1,0)$  & $3P_1-6P_2+3P_3$\\
$(2,1,1,1)$  & $4P_2-3P_3$\\
$(2,1,0,-1)$ & $P_3$\\
\bottomrule
\end{tabular}
\end{table}

For a selected edge $e$, let $d(e)$ count internal vertices on the left
spine starting at the old subtree root. Let $a(e)$ count consecutive
left-child edges from that root towards the tree root, including $e$
when it is a left-child edge. Define the threshold counts
\[
 N(r,s)=\#\{e:d(e)\geq r,\ a(e)\geq s\}.
\]
For $r,s\geq0$ and $r+s>0$, moving an edge $s$ left edges towards the root
gives a bijection with occurrences counted by $P_{r+s}$; descending those
edges gives its inverse. Hence $N(r,s)=P_{r+s}$, with $N(0,0)=E$.

For example, placing the old subtree on the right creates a branching
vertex of left-spine depth one. If $d(e)\geq2$ and $a(e)\geq1$, its
parent's depth falls from $d(e)+1\geq3$ to two; more distant left ancestors
remain at depth at least three. Exactly one $P_3$ occurrence is lost,
$P_2$ is unchanged, and two edges and one internal vertex are added.
This gives the last jump in Table~\ref{tab:remy-events}, $(2,1,0,-1)$,
with activity $N(2,1)=P_3$: each eligible edge admits one right placement.
The other classes follow from differences of adjacent threshold counts
(two-dimensional M\"obius inversion); grouping classes with identical jumps
gives Table~\ref{tab:remy-events}.

Summing the jumps times their activities in Table~\ref{tab:remy-events}
gives the four expectation equations in
Section~\mainref{sec:remy} of the main text.
The total activity is $2E$. The EMGF operator uses one term
$\kappa(e^{\lambda\cdot\delta_e}-1)a_e(\partial_\lambda)$ for each event.
Replacing its factor $e^{\lambda\cdot\delta_e}-1$ by
$e^{\lambda\cdot\delta_e}$ and setting $\kappa=1$ gives Behr's
weighted-growth operator, agreeing exactly with
Equation~(57g)~\cite{Behr2021StochasticCombinatorics}.
The identity holds for trees of arbitrary finite size. The
\href{\SourceRoot/docs/remy-trees.md}{R\'emy source guide} records the
separate concrete-tree checks.
 \FloatBarrier
\subsection{Eight Sites}
\label{app:eight-slots}
Each site has a uniquely owned $P,L$ pair and an optional \texttt{bound}
edge. The supplied axioms require eight sites in a simple undirected
\texttt{next} graph, with degree exactly two. Such a graph is a union of
cycles; the possible component lengths are $3,4,5,8$, since a remainder
of one or two sites cannot form a simple cycle. A path match containing
an edge has a unique continuation around its component. A single-site
match has two orientations, requiring a factor $1/2$ when expressed by
oriented cycle counts. Newly added sites have unspecified occupancy.

Let $H_{m,S}$ count oriented matches of a complete $m$-cycle whose sites
in $S$ must be occupied, with all other occupancies unrestricted. For
required occupied positions $S$ and disjoint required free positions $F$,
the missing-edge tests give the inclusion--exclusion identity
\[
 \sum_{T\subseteq F}(-1)^{|T|}H_{m,S\cup T}.
\]
The backend adds the admissible cycle completions and eliminates their
formal coefficients exactly. Unsupported disconnected or partially owned
patterns remain unresolved. This calculation uses no occupancy states.
The subsequent rank check finds dimension 47 among the 48 generated
functions; it does not participate in synthesis. Additional ownerless
$P,L$ vertices affect neither the anchored counts nor their transitions. The
\href{\SourceRoot/docs/eight-slot-topologies.md}{eight-site derivation}
gives the count arguments and recorded checks.
 \FloatBarrier
\subsection{DNA Walker}
\label{app:dna-walker}

The four rules are taken from the two-legged DNA walker of Danos et al.
\cite[Section 1.1]{DanosHeindelHonoratoZimmerStucki2020}. We use one walker
on a simple directed cycle of arbitrary finite length at least five.
Its two leg edges attach either to the same site or to adjacent sites.
Replacing each directed edge by an incidence vertex retains its identity,
including the two parallel legs of a compressed walker. Source and target
ports retain direction. Every deleted incidence vertex has exactly its two
matched ports; hence no unmatched incident edge obstructs the DPO deletion.

The counting procedure reduces a complete two-leg walker with connected
DNA-path context of fewer than five sites. At a fixed walker position,
that context extends uniquely along every permitted cycle. The reduced
pattern keeps the walker and its attachment sites. An additional walker,
excess legs, invalid incidence degrees or a shorter DNA cycle give zero;
other pattern forms remain unsupported. Starting with only the extended pattern, composition introduces the
compressed pattern and closes without enumerating scaffold states.

Let $E$ count the extended configuration and let $C_{\rm raw}$ count
injective matches of the compressed pattern. The parallel legs give two
maps per compressed configuration, so its indicator is $C=C_{\rm raw}/2$.
The raw incidence comparison uses unit per-match multipliers.
For the published configuration-rate convention,
the extension rules use multiplier $1/2$ and the compression rules use
multiplier $1$. Writing $e=\mathbb E[E]$ and $c=\mathbb E[C]$, the generated
normalized equations are
\[
 \dot e=(k_{F,E}+k_{B,E})c-(k_{F,C}+k_{B,C})e,
 \qquad \dot c=-\dot e,\qquad c+e=1.
\]
Here $F,B$ denote forward and backward leg movement, and the second
subscript denotes the resulting extended or compressed configuration.
Raw incidence counts and normalized configuration counts are compared
separately in the recorded results.

Each leg move changes the walker's centre by half a site, with its sign
read from the directed scaffold. Solving the generated equations at
stationarity gives
\[
 V=\frac{k_{F,E}k_{F,C}-k_{B,E}k_{B,C}}
          {k_{F,E}+k_{B,E}+k_{F,C}+k_{B,C}},
\]
for a positive denominator. This is symbolically equal to the published
mean-velocity formula. The
\href{\MaterialsURL}{material guide} links the separate concrete-host checks
and comparisons with the published equations.
 \FloatBarrier
\subsection{Molecular ROP}
\label{app:molecular-rop}

The constructor family contains one benzyl-initiated unbranched chain,
finite free-lactone pools, persistent feed markers and an optional two-state
Pd--N gate. Hydrogens are explicit vertices. Composition and reduction
produce the molecular patterns; sequence names such as $S_{AB}$ are assigned
afterwards.

Each recognised whole-repeat or end pattern has one oriented injective map
per occurrence. Free-lactone and chain-end patterns match different
components, so their product counts all pairs of maps without assuming
probabilistic independence. Every insertion selects the unique hydroxyl
end and one free lactone. Different copies give the same coordinate jump;
the Pd--N absence predicate is the gate indicator. These identities yield
the A/B-path events of Appendix~\ref{app:path-equations} under the molecular
count correspondence. The backend checks these rule forms and refuses
molecular vertex deletion or unsupported positive pattern forms.

\paragraph{Discarding impossible overlaps.}
The constructor family bounds incident bonds by bond type and neighbouring
atom label, as in the forbidden bond patterns used for aldol
addition~\cite[Definition~19 and Example~7]{BehrKrivineAndersenMerkle2021}.
Later identifications may merge neighbours that are still distinct in a
partial overlap. For example, identifying the oxygens of two O--H bonds
appears to give two hydrogen neighbours, but identifying their hydrogens
and bonds as well leaves one. Rejecting the partial overlap solely for
exceeding a one-hydrogen bound would lose that completion.
An entire search branch is therefore rejected only if no remaining
identifications can satisfy the bounds. An upper bound on the bonds that
could still become shared makes this pruning conservative; final
composition checks conditions and retains multiplicities.

\paragraph{Finite witnesses for zero counts.}
A positive embedding is an injective map preserving labels and edges,
without testing the pattern's condition. Matching is non-induced:
additional edges between mapped host vertices are allowed. If no positive
embedding exists, no additional condition can make the count nonzero.

For this ROP family, let $P$ be a nonempty connected labelled pattern with
$n$ vertices, and set
$h=\lfloor(n-1)/6\rfloor+2$. Let $\mathcal H_h$ contain the explicit-H
benzyl-initiated chains for every word over the selected alphabet of length
at most $h$, including the empty chain, together with one representative of
each free-lactone and feed-marker type and both gate states when present.
If $P$ has no positive injective embedding into any member of $\mathcal H_h$,
then $[P,c](G)=0$ for every constructor state $G\in\Omega$ and every
additional application condition $c$.

To see why, a connected image lies in one host component. All component types other
than the polymer already occur in $\mathcal H_h$. In a polymer, the repeat
backbone is a path; carbonyl oxygens and carbon-bound hydrogens do not
provide shortcuts. If an image meets $t$ consecutive repeat blocks, a
simple path in the image between the first and last traverses every
intermediate block. Each such block contributes at least six backbone
edges (seven for $B$), so $6(t-2)\leq n-1$ and $t\leq h$.
Keep the interval of blocks met by the image and remove untouched outer
blocks. The benzyl initiator and terminal hydrogen can be attached outside
the image; when the image meets either original end, keep that end.
This preserves every mapped vertex label and positive edge, including
explicit hydrogens. Additional edges outside the pattern are allowed by
non-induced matching. The resulting positive embedding is into a chain
listed in $\mathcal H_h$. Images contained entirely in the initiator are
covered by the empty chain. Thus a positive embedding in any constructor
state would yield one in the finite cover. Any match satisfying $c$ also
requires such a positive embedding.

Failure of every embedding check certifies zero. A successful embedding
does not determine the count: recognised repeat, pool and end patterns use
the identities above; other positive patterns remain unsupported. The
implementation tests connected components and refuses covers beyond its
budget, rather than truncating a cover and declaring zero. The
\href{\SourceRoot/docs/molecular-rop-discovery.md}{molecular backend guide}
records these limits and the independent concrete-host checks.
 \FloatBarrier
\subsection{Chlorination with a Neighbour NAC}
\label{app:chlorination-equations}

The ring order is $0,1,2,3,4,0$. Site 0 is the charged, tail-bearing anchor;
neutral sites $1,\ldots,4$ bear H or Cl. The comparison removes only the
NAC, leaving both rules and the graph family otherwise unchanged. Each reaction replaces one C--H
bond by C--Cl, consumes one HCl and releases one H$_2$. The NAC forbids a
neighbouring neutral ring carbon bearing chlorine. With fixed total chlorine
$H_0=4$, the HCl inventory is $h=4-Z$. Each enabled site has $h$ injective
reaction matches, hence rate $\kappa h$. Growth preserves the ring and
HCl inventory and has zero jump on all discovered coordinates.

Write $h_i\in\{0,1\}$ for the presence of H at site $i$. The positive
pattern counts discovered from the single query $Z$ have the identities
\begin{equation}
 \begin{aligned}
 Z&=4-\sum_{i=1}^4h_i,& P_1&=h_1h_2+h_3h_4,\\
 P_2&=h_1h_2h_3+h_2h_3h_4,& P_3&=2h_1h_2h_3h_4.
 \end{aligned}
 \label{eq:chlor-nac-counts}
\end{equation}
The factor two in $P_3$ retains the two embeddings related by reflection.
Unspecified substituents in these positive patterns are unconstrained:
these are not indicators of fully specified substitution states.

The backend reduces the actual transported condition expressions to
Boolean polynomials, using $h_i^2=h_i$, and expresses them as positive
pattern counts. The support worklist constructs these patterns from the
conditional-count monomials; no target basis or ring-state transition
table is supplied. The normalized chlorination generator images are
\begin{equation}
 \begin{aligned}
 \Gen_{\rm chlor}Z&=(4-Z)(P_1+P_2),&
 \Gen_{\rm chlor}P_1&=-(4-Z)(P_1+P_2),\\
 \Gen_{\rm chlor}P_2&=-(4-Z)(2P_2+P_3),&
 \Gen_{\rm chlor}P_3&=-4(4-Z)P_3.
 \end{aligned}
 \label{eq:chlor-nac-images}
\end{equation}
These are function identities, not closed ODEs obtained by substituting
mean counts into products.

Partitioning symbolic enabled source matches by their full jump vectors
gives the following events, with coordinates ordered as $(Z,P_1,P_2,P_3)$.
The rate of each event is $\kappa$ times its activity.
\begin{center}
\begin{tabular}{ll}
\toprule
Jump & Activity\\
\midrule
$(1,-1,-2,-2)$ & $(4-Z)P_3$\\
$(1,-1,-1,-2)$ & $(4-Z)P_3$\\
$(1,-1,-1,0)$ & $2(4-Z)(P_2-P_3)$\\
$(1,-1,0,0)$ & $(4-Z)(P_1-P_2)$\\
\bottomrule
\end{tabular}
\end{center}
For the last event, site 1 reacts when $h_1=h_2=1$ and $h_3=0$,
or site 4 reacts in the reflected configuration. Each destroys one
$P_1$ embedding and leaves $P_2,P_3$ unchanged. The number of eligible sites is
\[
 h_1h_2(1-h_3)+h_3h_4(1-h_2)=P_1-P_2.
\]
Multiplication by the $4-Z$ choices of HCl gives the stated activity.
Partitioning all enabled matches this way gives disjoint, exhaustive
jump classes. Different matches in one host may have different jumps;
this construction does not invoke Algorithm~\mainref{alg:main}'s
uniform-match step. Without the NAC, the single event on $Z$ has jump $+1$
and activity $(4-Z)^2$.

On this family, aromatic bonds and C--Cl bonds occur only in the ring.
Growth acts on a terminal carbon with three H neighbours and methane
with four; each neutral ring carbon has at most one. Its bond changes
therefore preserve the ring, its H/Cl substituents and the HCl inventory,
and create no new aromatic or C--Cl contexts. The count identities thus
hold for arbitrary finite tails and finite methane/hydrogen pools.
The \href{\ChlorinationResultsURL}{reproduction profile} generates the
condition reductions and separate joint-rate checks against public
MolFoundry applications and independent site updates.

\paragraph{Forward equations and chlorination arrest.}

Initially the ring is unchlorinated and four HCl molecules are present.
The guarded initial coordinates are $(0,2,2,2)$. Only after event synthesis
do we explore their reachable coordinate states:
\[
 (0,2,2,2),\quad(1,1,0,0),\quad(1,1,1,0),\quad(2,0,0,0).
\]
In this order the rate matrix, with source states as rows, is
\begin{equation}
 Q_{\rm NAC}=\kappa
 \begin{pmatrix}
 -16&8&8&0\\0&-3&0&3\\0&0&-6&6\\0&0&0&0
 \end{pmatrix}.
 \label{eq:chlor-nac-Q}
\end{equation}
Without the NAC the reachable coordinates are $Z=0,1,2,3,4$, with successive
rates $\kappa(16,9,4,1)$.
For either matrix, $p'(t)=Q^{\mathsf T}p(t)$ and
$p(0)=(1,0,\ldots,0)^{\mathsf T}$.
These are finite forward equations of the generated event models, not a
claimed output of linear expectation closure.

Writing $p=(p_0,p_1,p_2,p_3)^{\mathsf T}$ in the guarded state order,
\[
 \mathbb E[Z]=p_1+p_2+2p_3,\qquad
 \Pr(\text{blocked with free sites remaining})=p_3.
\]
The only reachable absorbing state has $Z=2$, and every other state has
positive exit rate when $\kappa>0$. Chlorination therefore stops at two
chlorines with probability one. Without the NAC, the successive positive
rates lead to $Z=4$; no blocked state with free sites remains reachable.
These forward equations give the curves in the main text. Tail growth
can still occur in the full graph process.
 \FloatBarrier
\clearpage
\section{Composition Benchmark}
\label{app:composition-benchmark}

The query is the directed path $ABAB\cdots$ on $n$ vertices. Six condition-free rules
replace $p\to T$ by $p\to s\to T$, for $p\in\{R,A,B\}$ and
$s\in\{A,B\}$, preserving $p$ and $T$. Each rule is composed with the
query identity in both orders. No model vertex is deleted, so DPO and SPO
admissibility coincide here. FRAGGER uses directed multigraphs; MolFoundry
uses their incidence encoding, with $2n-1$ vertices for the query and no
path-family pruning.

Runs are serial on one native Linux x86\_64 host (Xeon w7-2595X).
MolFoundry~0.1.9 (14 October 2026) uses Python~3.12.3; FRAGGER revision \texttt{0d9595c}
(9 October 2026) uses Scala~2.13.2 and Java~11 with \texttt{-Xmx4g}. Each tool/size uses a fresh process: four small global warm-ups,
two size-specific warm-ups, then five timed repetitions (reported as medians). Startup, compilation and input
construction are excluded. MolFoundry timings include canonical grouping
and representative GML serialization; FRAGGER timings cover native
left/right-union enumeration. These timings measure composition,
not complete closure analysis.

A separate MolFoundry-only probe with the same protocol gives
$0.927$, $2.14$, $7.79$ and $21.2$ seconds at $n=48,64,96,128$,
respectively. The
\href{\SourceRoot/scripts/comparison/ab_scaling/README.md}{benchmark guide}
gives the commands, medians and recorded environment.
 
\FloatBarrier
\paragraph{AI assistance.}
OpenAI's ChatGPT assisted with exploratory prototyping and testing of the
accompanying software. The authors reviewed the resulting contributions and
retain responsibility for the implementation and reported results.
 \end{document}